\documentclass[journal=nalefd,manuscript=letter,layout=traditional,super=true,]{achemso}
\usepackage[version=3]{mhchem} 

\title{Purcell enhanced and indistinguishable single-photon generation from quantum dots coupled to on-chip integrated ring resonators}

\author{\L{}ukasz Dusanowski}
\email{lukasz.dusanowski@uni-wuerzburg.de}
\affiliation{Technische Physik and W\"{u}rzburg-Dresden Cluster of Excellence ct.qmat, University of W\"{u}rzburg, Physikalisches Institut and Wilhelm-Conrad-R\"{o}ntgen-Research Center for Complex Material Systems, Am Hubland, D-97074 W\"{u}rzburg, Germany}

\author{Dominik K\"{o}ck}
\affiliation{Technische Physik and W\"{u}rzburg-Dresden Cluster of Excellence ct.qmat, University of W\"{u}rzburg, Physikalisches Institut and Wilhelm-Conrad-R\"{o}ntgen-Research Center for Complex Material Systems, Am Hubland, D-97074 W\"{u}rzburg, Germany}

\author{Eunso Shin}
\affiliation{Department of Physics, Chung-Ang University, 156-756 Seoul, Korea}

\author{Soon-Hong Kwon}
\affiliation{Department of Physics, Chung-Ang University, 156-756 Seoul, Korea}

\author{Christian Schneider}
\affiliation{Technische Physik and W\"{u}rzburg-Dresden Cluster of Excellence ct.qmat, University of W\"{u}rzburg, Physikalisches Institut and Wilhelm-Conrad-R\"{o}ntgen-Research Center for Complex Material Systems, Am Hubland, D-97074 W\"{u}rzburg, Germany}
\altaffiliation{Institute of Physics, University of Oldenburg, D-26129 Oldenburg, Germany}

\author{Sven H\"{o}fling}
\affiliation{Technische Physik and W\"{u}rzburg-Dresden Cluster of Excellence ct.qmat, University of W\"{u}rzburg, Physikalisches Institut and Wilhelm-Conrad-R\"{o}ntgen-Research Center for Complex Material Systems, Am Hubland, D-97074 W\"{u}rzburg, Germany}
\altaffiliation{SUPA, School of Physics and Astronomy, University of St Andrews, KY16 9SS St Andrews, UK}

\date{\today}


\begin{document}
	
	\begin{abstract}
		{Integrated photonic circuits provide a versatile toolbox of functionalities for advanced quantum optics applications. Here, we demonstrate an essential component of such a system in the form of a Purcell enhanced single-photon source based on a quantum dot coupled to a robust on-chip integrated resonator. For that, we develop GaAs monolithic ring cavities based on distributed Bragg reflector ridge waveguides. Under resonant excitation conditions, we observe an over twofold spontaneous emission rate enhancement using Purcell effect and gain a full coherent optical control of a QD-two-level system via Rabi oscillations. Furthermore, we demonstrate an on-demand single-photon generation with strongly suppressed multi-photon emission probability as low as 1\% and two-photon interference with visibility up to 95\%. This integrated single-photon source can be readily scaled up, promising a realistic pathway for scalable on-chip linear optical quantum simulation, quantum computation, and quantum networks.}  
	\end{abstract}
	
	\maketitle
	Advanced quantum optics applications such as quantum networks, quantum simulation and quantum computing require single-photon sources with simultaneously high efficiency and degree of photons indistinguishability. Among different kinds of emitters, self-assembled quantum dots (QDs) coupled to optical cavities have been shown to be one of the brightest on-demand single-photon sources up to date (SPS)~\cite{Santori2002,Aharonovich2016,Senellart2017}, which can simultaneously reach almost unity single-photon indistinguishability and purity as well extraction efficiency as high as 60\%~\cite{Somaschi2016,Unsleber2016,Senellart2017,Wang2019}. Those advances allowed already for demonstration of on-demand CNOT-gates,~\cite{Pooley2012,Gazzano2013a,He2013a} heralded entanglement between distant spin qubits,~\cite{Delteil2015,Stockill2017} quantum teleportation,~\cite{Nilsson2013,Reindl2018} or the recent realization of 20-photon boson sampling~\cite{Wang2019-sampling}. This tremendous progress in QD-based single-photon sources has been achieved by a combination of resonant excitation~\cite{Somaschi2016,Unsleber2016,Senellart2017,Wang2019,He2013a,Wang2019-sampling,Flagg2009,Ates2009,NickVamivakas2009} to eliminate emission time jitter and cavity quantum electrodynamics to overcome fundamental limitations set by intrinsic exciton-phonon scattering inherent in solid-state platform\cite{Iles-Smith2017}. 
	
	As the quantum optics experiments have increased in complexity, there has been an increasing need to transition bulk optics experiments to integrated photonic platforms, where all components could be placed on the same chip. It means that single photon sources, photonic circuitry, and detectors could be optically interconnected with very small losses and allow control of single-photon states with greater fidelity. More importantly, integrated circuits combined with quantum emitters are believed to be a reliable approach to achieve full scalability toward large scale quantum optics.~\cite{Dietrich2016a,Aharonovich2016,Hepp2019}
	
	The GaAs material system with embedded QDs seems to be a perfectly suited system for that purpose, where both fully homogeneous~\cite{Jons2015,Enderlin2012,Schwagmann2011,Arcari2014,Reithmaier2015} and heterogeneous~\cite{Davanco2017,Elshaari2017,Kim2017,Ellis2018} integrated photonic circuits were realized. In this approach, light can be directly coupled into in-plane waveguides (WGs) and combined with other functionalities on a chip such as phase shifters~\cite{Midolo2017,Wang2014a}, beam splitters~\cite{Prtljaga2014,Jons2015,Kim2017}, filters~\cite{Elshaari2017,Elshaari2018,Aghaeimeibodi2019}, detectors~\cite{Reithmaier2015,Kaniber2016} and other devices for light propagation, manipulation and detection on a single photon level. Moreover, integrated circuits allow for spatial separation of excitation and detection spots, which straightforwardly enables applying resonant driving schemes to slow-down decoherence processes and reduce on-demand emission time-jitter.~\cite{Makhonin2014,Kalliakos2016,Schwartz2016,Kirsanske2017b,Liu2018,Dusanowski2019} 
	
	Among different on-chip QD-integration implementations, a particular emphasis over the last few years has been devoted to photonic crystal (PhC) and nanobeam WGs~\cite{Lund-Hansen2008,Enderlin2012,Arcari2014,Stepanov2015}. In particular, pulsed resonance fluorescence generation of indistinguishable single photons in waveguides was achieved recently.~\cite{Kirsanske2017b,Liu2018} Despite this progress, the persistent problems of PhC and nanobeam WGs systems are their large propagation losses and fragility,~\cite{Hepp2019} limiting dimensions of freestanding circuits to a few hundred microns. A combination of PhC and nanobeam WGs with GaAs ridge waveguides\cite{Fattahpoor2013} or heterogeneous integration with SiO$_2$/Si$_3$N$_4$ ridge waveguides~\cite{Davanco2017,Elshaari2017,Kim2017} has been also proposed; however, it demands complicated multi-step fabrication process and introduces losses at interfaces. An alternative approach offering more robust design with high mechanical stability is monolithic ridge waveguides. Typically, those are based on GaAs core layer and distributed Bragg reflector (DBR) or AlGaAs layer cladding. Using this system, complex integrated circuits with large footprints have been realized\cite{Wang2014a,Jons2015,Schwartz2018} promising a clear path toward scalability. Using GaAs ridge waveguides, full on-chip second-order correlation experiments have been implemented\cite{Schwartz2018} and pulsed resonance fluorescence has been demonstrated,\cite{Schwartz2016,Reigue2017,Dusanowski2019} which allowed recently for single-photon generation with non-corrected two-photon interference visibility of 97.5\%\cite{Dusanowski2019}. The main challenge in the broader application of GaAs ridge waveguides with QDs in quantum integrated photonics is relatively low coupling efficiency of photons emitted from QD to waveguide mode (15-22\% into each WG arm) due to the low-refractive-index contrast between GaAs and AlGaAs materials. One route to overcome this problem is the application of the Purcell effect to boost up QD emission funneling into the WG mode\cite{Hepp2018}.
	
	Hereby we will combine cavities with ridge waveguides and show that such a system has the potential of achieving simultaneously high coupling efficiency and near-unity single-photon indistinguishability in reliable integrated circuit platform. Among different types of cavities, ring resonators are well established on-chip functionalities in integrated circuits\cite{Rabus2007}. They are commonly used as filters\cite{Davanco2017,Elshaari2017,Elshaari2018}, switches\cite{Wen2011} or parametric down-conversion pair sources\cite{Engin2013}. Because of easily achievable high-quality factors (Q), they are also perfectly suited for increasing light-matter interaction with quantum emitters\cite{Davanco2017,Faraon2011}. Recently, devices combining the InAs QDs with ring resonators have been realized by the heterogeneous integration of GaAs and Si$_3$N$_4$ platforms. Spontaneous emission enhancement factors up to 4 have been observed\cite{Davanco2017}, however, experiments have been limited to non-resonant continuous-wave (cw) excitation only, which is known to degrade the purity and indistinguishability of emitted photons, so that high-performance on-demand SPS ring devices were not realized yet.
	
	In this article, we demonstrate resonantly driven triggered SPSs consisting of self-assembled InAs/GaAs QDs coupled to on-chip ring resonators based on the DBR ridge waveguides within the same GaAs wafer. By spectral tuning of the QD emission into the ring cavity mode, we achieve an over twofold spontaneous emission rate enhancement, revealed by time-resolved studies. Furthermore, we observe Rabi oscillations visible from the QD emission intensity variation as a function of pump pulse area, demonstrating coherent optical control. Finally, we show a generation of single and indistinguishable photons on demand by performing second-order correlation experiments in Hanbury-Brown-Twiss (HBT) and Hong-Ou-Mandel (HOM) configurations. By employing resonant excitation scheme within on-chip ring-bus-waveguide system we observe strongly suppressed multi-photon emission probability better than 1\% and two-photon interference with visibility up to 95\%. The combination of those results makes our DBR ridge ring resonator single-photon source competitive with other integrated structures approaches.  
	
	\begin{figure*}
		\includegraphics[width=6.5in]{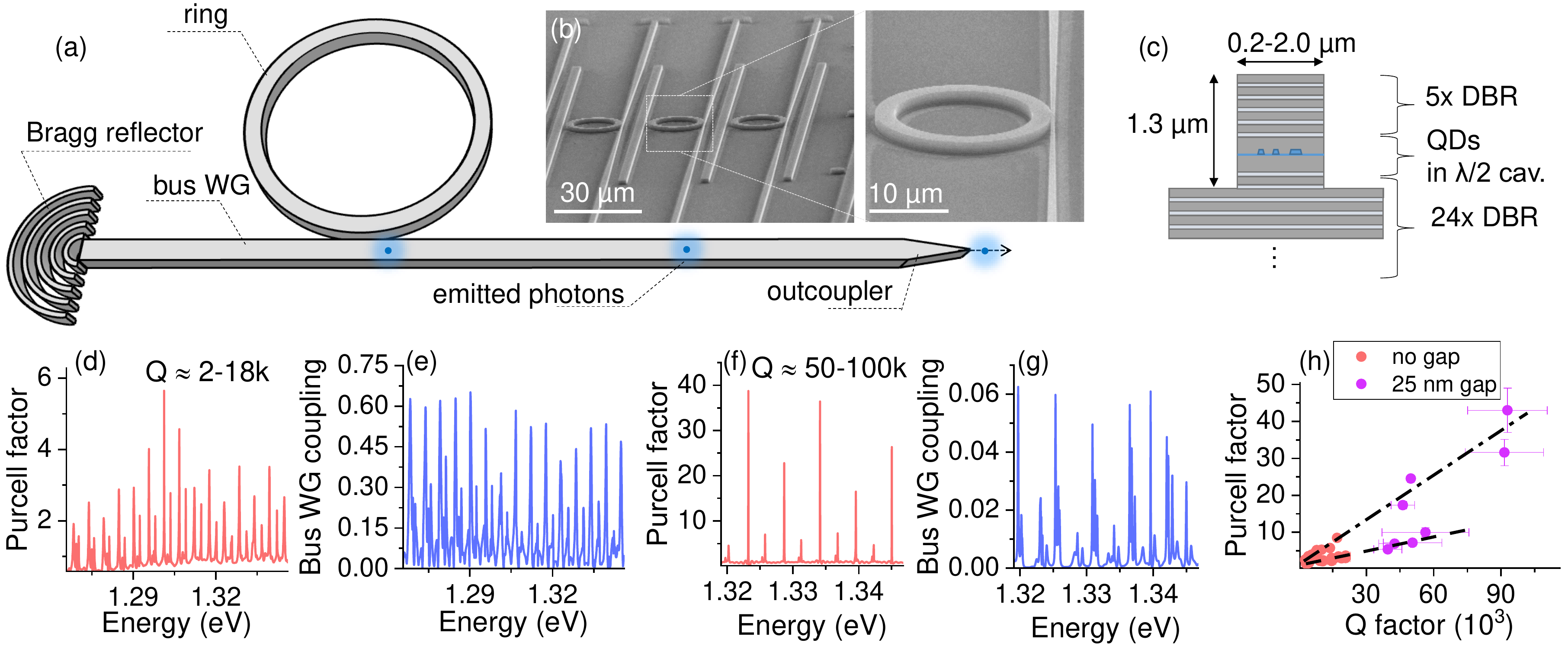}
		\caption{\label{fig:1}On-chip integrated ring resonator device. (a)~Artistic scheme of the DBR waveguide (WG) based ring resonator. Single quantum dots are placed within the core of the WG and excited optically from the top. Emitted photons are collected from the side facet of the structure within the tapered out-coupler. (b)~Scanning electron microscope images of the fabricated ring resonator devices with radius $R$ of 10~$\mu$m. (c)~DBR WG cross-section with marked layers. (d),(f)~Simulated Purcell factor vs energy for the 2~$\mu$m width ring resonator with an outer radius of 10~$\mu$m coupled to 0.2~$\mu$m width bus WG and 0 and 25~nm ring-bus-WG gap, respectively. (e),(g)~Simulated QD emission coupling efficiency into the bus WG for 0 and 25~nm gap structures, respectively. Very high quality factor Q of the 25~nm gap ring cavity required limiting the simulation spectral window to 20~nm. (h)~Purcell factor vs Q factor taken from panels d and f, revealing clear linear dependence for fundamental (dot-dashed line) and higher-order radial modes (dashed line).
		}
	\end{figure*}
	
	An artistic sketch of our integrated system is shown in Fig.\ref{fig:1}(a). It consists of the following elements: the ring resonator coupled in-plane to bus waveguide, circular Bragg grating on the one end of the WG working as a reflector and tapered out-coupler designed to minimize reflection and optimize out-coupling efficiency into the off-chip collection optics (more details in Supporting Information). To operate our device as an on-chip SPS, the QD is excited from the top of a ring and the single-photon emission is collected from the bus waveguide after more than 1~mm travel distance by the tapered out-coupler localized on the sample edge. The cross-section of the DBR WG is shown in Fig.\ref{fig:1}(c).
	
	In this work, we consider two ring resonator designs: (i)~based on uniform single-mode (SM) WG width of 0.8~$\mu$m of both the ring and the bus section and (ii)~based on multi-mode (MM) 2.0~$\mu$m width WG uniform within the ring and tapered bus WG narrowing from 2.0~$\mu$m to single-mode 0.2~$\mu$m width section close to the coupling region. A tapering section in case of second design ensures that light is coupled and guided only in the fundamental transverse-electric (TE) mode of the MM bus waveguide, which can be latter efficiently back coupled to single-mode WG using mode converters (more details in Supporting Information). The first design is simpler and more straightforward to combine with other on-chip SM functionalities, such as beam-splitters or interferometers. However, small width WGs tend to have larger propagation losses, since a less tightly confined mode is more susceptible to scattering due to sidewalls roughness~\cite{KuanPeiYap2009}. Indeed, in case of the considered system, we observe propagation losses on the level of 2-3.5~dB/mm for 2.0~$\mu$m width WGs and 4-7 dB/mm in the case of 0.8~$\mu$m WGs.
	
	To optimize the structure geometry for maximized Purcell, we fabricated a set of ring devices with various diameter and ring-bus-WG spacing (gap), and performed a systematic check of the Q-factors via optical measurements. For rings with $R \leq 5~\mu$m the quality factor was limited to 1-2k due to the fabrication imperfections (mainly surface roughness). On the contrary, for rings with $R \geq10~\mu$m modes with Q exceeding 7-12k were observed. Within such mode volume to Q factor trade-off, we concluded that rings with $R$~=~10~$\mu$m are the most promising in terms of obtainable Purcell (more details in Supporting Information). 
	
	For the quantitative estimate of the maximal obtainable Purcell enhancement and waveguide coupling efficiency, we simulated photonic properties of our devices using Finite Difference Time Domain (FDTD) method. By placing a point dipole in the ring cavity mode, we calculated Purcell enhancement factor, defined as power emitted by a dipole source in the ring normalized to the power emitted by the dipole in a homogeneous (bulk) environment. Results of simulations for a ring with WG profile of 2.0~x~1.3~$\mu$m, 10~$\mu$m outer radius and 0.2~$\mu$m bus WG are presented in Figures~\ref{fig:1}(d) and (f) for gap-less and 25~nm gap structures, respectively. Clear sharp peaks at energies fulfilling traveling wave resonance conditions $\lambda m = 2 \pi n_g R$ for fundamental and higher-order radial modes can be distinguished, where $m$ is a mode number, $\lambda$ is a wavelength, $n_g$ is a group refractive index and $R$ is a ring radius. The simulated Purcell factors can reach values up to 6 with Q of 2k-18k for gap-less rings and Purcell up to 40 with Q of 50-100k for 25~nm gap rings. In Fig.\ref{fig:1}(e) the efficiency of the QD emission coupling into the bus WG is plotted for gap-less ring device. Values as high as 67\% are obtainable, which consist of 70\% QD-to-ring and 96\% ring-to-bus-WG coupling contributions. Despite very high Purcell for 25~nm gap rings and thus almost unity coupling into the ring cavity mode, a relatively low coupling of 2-6\% into the bus WG is expected, as shown in Fig.~\ref{fig:1}(g). With successive increase of the gap to 50~nm, WG coupling decreases even further, while Purcell seems to be unaffected (see Supporting Information). Taken together, these results suggest that maximal obtainable Purcell for 10~$\mu$m radius rings is 40 and it is limited by bending losses. By plotting the Purcell factor vs Q for each mode, linear dependencies for fundamental and higher-order radial modes are observed as shown in Fig.~\ref{fig:1}(h). Following this relation, Purcell of 5-6 for Q of 10k and 10-12 for Q of 20k is expected. Similar simulations for rings placed inside the bus WG (negative gap) are summarized in Supporting Information. Interestingly, we found that for -100~nm gap, the total bus WG coupling efficiencies of up to 90\% seem to be feasible with a moderate Purcell of around 2.   
	
\begin{figure}
		\includegraphics[width=3.2in]{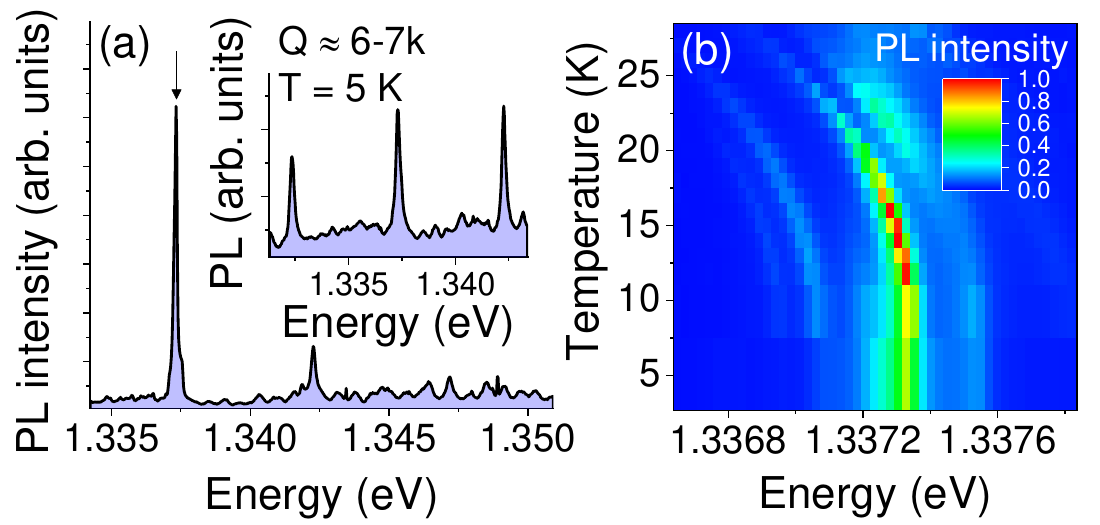}
		\caption{\label{fig:2} (a)~Side-detected photoluminescence spectrum from QD$_1$ recorded under non-resonant cw excitation and temperature of 5~K. Inset:~optical modes of the ring under investigation. (b)~Temperature dependence of the QD$_1$ PL spectrum.
		}
	\end{figure}

	For optical characterization of our ring devices, we use top-excitation and side-detection micro-photoluminescence setup and cool down the sample to 4.5~K using a high-stability closed-cycle cryostat. The PL signal is collected from side facet after around 1~mm travel distance in the bus WG. First, we consider MM ring resonator with 10~$\mu$m radius and 100~nm nominal gap between the ring and tapered bus WG. In Figure~\ref{fig:1}(b) scanning electron microscope (SEM) images of the example fabricated ring resonator are shown. Higher magnification SEM images suggest that the GaAs cavity layer is not completely etched (effectively gap is $\textless$100~nm). Figure~\ref{fig:2}(a) shows a side collected photoluminescence (PL) spectra from a QD$_1$ under above-band gap cw excitation. A single emission line at 1.337~eV is visible with the setup resolution limited linewidth. The inset in Fig.~\ref{fig:2}(a) shows investigated ring modes with Q factor values of around 6-7k, probed by photoluminescence at high power (the ensemble of QDs acts as a spectrally broad light source). We tune the QD emission energy across the ring cavity mode by increasing the sample temperature. At around 14~K, a resonance of the QD emission with the ring mode is established, as shown in Fig.~\ref{fig:2}(b). 
	
\begin{figure*}
		\includegraphics[width=6.1in]{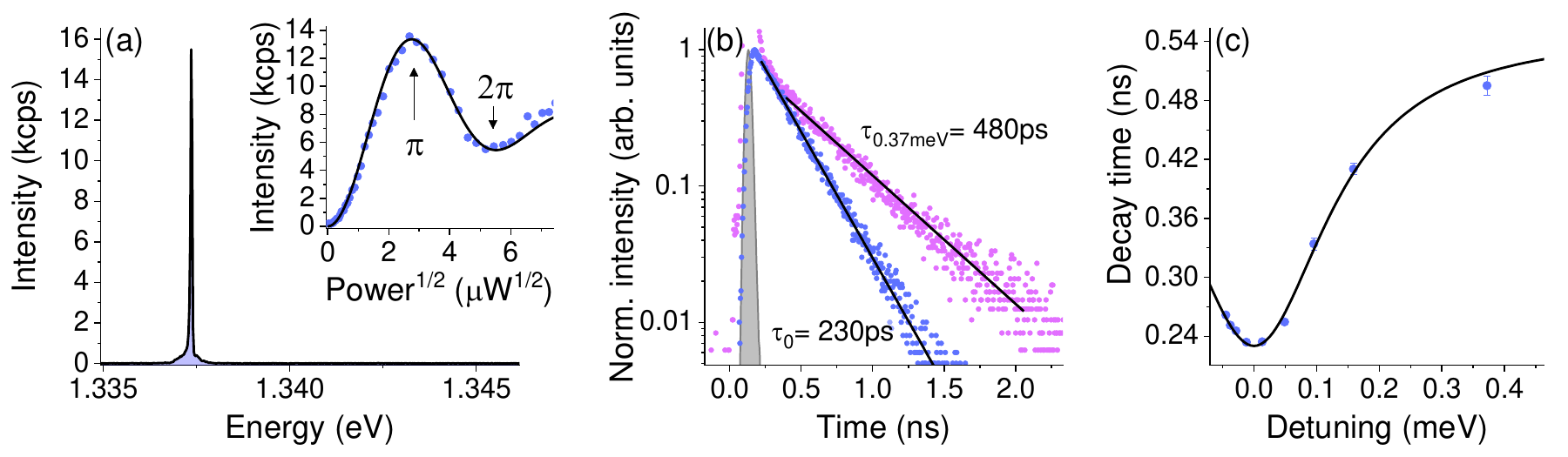}
		\caption{\label{fig:3} (a)~Pulsed resonance fluorescence spectrum of QD$_1$ at 5~K. Inset:~Resonance fluorescence intensity vs square root of power demonstrating Rabi oscillations. (b)~Time-resolved resonance fluorescence traces recorded for 0.37~meV and no QD-cavity detuning. (c)~Resonance fluorescence decay time vs QD-cavity energy detuning demonstrating Purcell enhancement.
		}
	\end{figure*}

	Next, device characteristics under s-shell resonant excitation were checked. Figure~\ref{fig:3}(a) shows side-detected pulsed resonance fluorescence spectra from QD$_1$. In the inset of Fig.~\ref{fig:3}(a) the peak intensity versus the square root of the incident power is shown. Clear oscillatory behavior with damping at higher power is observed, which is clear evidence of Rabi oscillations related to coherent control of the QD two-level system. The data have been fitted with exponentially damped cosine function assuming phonon-related dephasing included in quadratic exponential damping term\cite{Gerhardt2018}. 
	
	To verify Purcell enhancement, we performed time-resolved resonance fluorescence measurements for different QD-cavity detunings as plotted in Fig.~\ref{fig:3}(b)-(c). All time traces exhibit mono-exponential decays. In strongly detuned case, beside QD-related emission decay, non-filtered laser contribution is also visible. At a temperature of 14~K, corresponding to the QD-cavity resonance, the shortest lifetime of 230~ps has been observed [time trace with blue points in Fig.~\ref{fig:3}(b)]. In the case of 0.37~meV detuning, [magenta points in Fig.~\ref{fig:3}(b)] time constant of 480~ps has been recorded. For comparison, in the case of QDs coupled to straight ridge waveguides, we observe decay time constants on the level of 500-580~ps under resonant driving conditions\cite{Dusanowski2019}. By comparing 230~ps with the 500~ps lifetime, we can extract an over twofold radiative rate enhancement for the considered QD. The measured decay time as a function of detuning [Fig.~\ref{fig:3}(c)] is well fitted by the standard weak-coupling theoretical model (black curve) assuming coupling to the cavity mode with a quality factor of 7k, and considering that the emission rate into the cavity follows a Lorentzian dependence with respect to the detuning. From the fit, we extracted a maximum 2.4-fold enhancement of the spontaneous emission rate with Purcell factor $F_p$ of 1.4 assuming that the QD emission rate outside the ring cavity mode is equal to the emission rate in the bulk GaAs (more details in Supporting Information). For the Purcell of 1.4, coupling efficiency into the cavity mode $\beta$, reaches values as high as 58\% following the formula $\beta = F_P/(F_P+1)$ with respect to 30-44\% coupling efficiency limitation in the non-structured WGs. Assuming 96\% coupling between ring and bus WG, we expect an overall coupling efficiency of 55\%. To cross-check this value, we additionally estimated the QD-bus-WG coupling efficiency, based on the single-photon detector counts and total efficiency of the circuit (more details in Supporting Information). By careful calibration of the setup transmission, we obtained the total device extraction efficiency of 2.4\% into the first detection lens and lower limit of QD-WG coupling on the level of 10\%. We point out that, QD-WG coupling efficiency was derived assuming that all on-chip functionalities such as reflector, tapers and out-coupler perform as good as theoretically expected, which most likely strongly underestimates the calculated efficiency value (more details in Supporting Information).
	
\begin{figure*}
		\includegraphics[width=5.0in]{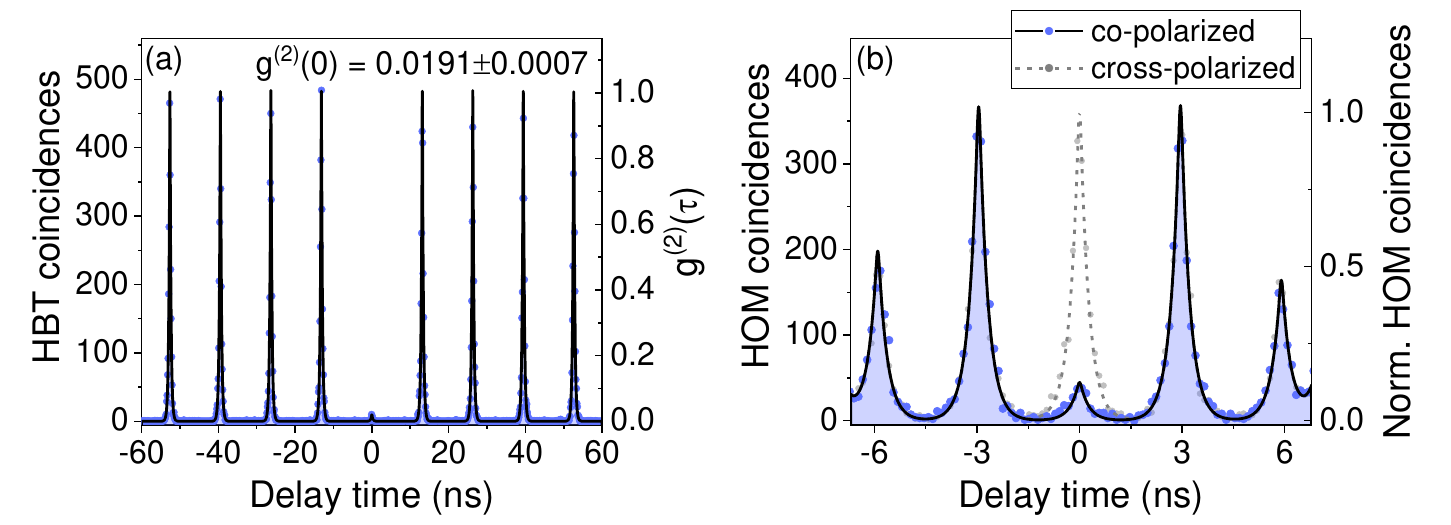}
		\caption{\label{fig:4} (a)~Second order correlation function histogram recorded for pulsed resonance fluorescence under $\pi$-pulse excitation power. The value of $g^{(2)}(0)$ is calculated from the integrated photon counts in the zero-delay peak divided by the average of the adjacent set of peaks, while the uncertainty of the $g^{(2)}(0)$ is based on the standard deviation of the Poissonian peaks integrated counts. (b)~Two-photon interference HOM histogram recorded for 3~ns time separated co- (blue pints) and cross-polarized (gray points) single photons under pulsed resonance fluorescence and $\pi$-pulse excitation power.	
		}
	\end{figure*}

	To characterize our ring device single-photon emission statistics, auto-correlation experiments at a temperature of 5~K have been performed on the resonance fluorescence signal filtered out from a broader laser profile and phonon sidebands. In Fig.~\ref{fig:4}(a), a second-order correlation function histogram recorded in HBT configuration under $\pi$-pulse excitation is shown. At zero delay, it shows clear antibunching with almost perfectly vanished multi-photon emission probability of $g^{(2)}(0)=0.0191\pm0.007$. The experimental data is fitted by two-sided mono-exponential decay with a time constant of 260~ps, convoluted with the setup instrumental response function (IRF) with a width of 50~ps.
	
	Next, we test the indistinguishability of emitted photons through HOM interference experiments. For that purpose, we excite QD$_1$ by a pair of pulses separated by 3~ns. Two subsequently emitted photons are then filtered by a monochromator and introduced into fiber-based 3~ns delay unbalanced interferometer, where a delay between them is compensated to superimpose single-photon pulses on the beam splitter~\cite{Santori2002}. The resultant two-photon interference histogram for orthogonal (gray points) and parallel (blue points) polarized photons is shown in Figure~\ref{fig:4}(b). The histogram consists of five 3~ns delayed peaks of the central cluster. In the case of identical polarizations, an almost vanishing zero-delay peak is observed. In contrast, for two photons with cross-polarization, the zero-delay peak has the same intensity as its adjacent $\pm$3~ns delayed peaks. To evaluate the zero-delay peak area in respect to the neighboring peaks, the experimental data have been fitted with the two-side exponential decay functions convoluted with the setup IRF (more details in Supporting Information). Using the fitting procedure, we obtain a raw value of two-photon interference visibility of 0.90$\pm$0.02. After including the residual multi-photon probability of $g^{(2)}(0)=0.0191$, as well as non-perfect interferometer visibility and splitting ratio~\cite{Santori2002} we determined a corrected degree of indistinguishability to be 0.95$\pm$0.02.
	
	Similar optical characterization has been repeated on two other ring devices with radius of 10~$\mu$m and 40~$\mu$m. In the case of second 10~$\mu$m radius ring, the shortest lifetime of 210~ps was observed ($F_p$~=~1.7) with $g^{(2)}(0)$ of $0.035\pm0.001$ and corrected indistinguishability of 0.93$\pm0.02$. In the case of 40~$\mu$m radius ring, no significant Purcell enhancement was observed (lifetime of 400~ps), while the emitter indicated very good single-photon performance with $g^{(2)}(0)$ of $0.04\pm0.01$ and corrected indistinguishability of 0.95$\pm$0.02 (corresponding graphs and more details in Supporting Information). 
	
	The demonstration of high-performance SPSs coupled to on-chip resonators is an important step toward large scale implementations of quantum photonic circuits. Within the investigated GaAs platform, our QD-ring devices could be straightforwardly combined with integrated beam splitters\cite{Wang2014a,Rengstl2015,Jons2015} which inherently offer near-perfect mode-overlap and very-high stability of the optical path-lengths, that enable high-fidelity quantum interference on-chip, that is, an essential component for quantum information processing. Monolithic integration of QDs with GaAs circuits might be thus considered as a reliable pathway toward full on-chip scalability, both in terms of realizing a mechanically stable large footprint circuit, as well as in terms of fulfilling 1\% two-photon gate operation error threshold and 67\% total efficiency threshold required for fault-tolerant quantum computing\cite{Varnava2008,Fowler2009}.       
	
	In this Letter, we have demonstrated the on-demand single-photon source based on InAs QD coupled to on-chip ring resonator. By combining a robust DBR ridge WG platform with the ring cavity, we overcome an intrinsic limitation of the monolithic ridge WG's coupling efficiency by utilizing a Purcell effect. We observed an over two-fold spontaneous emission enhancement, proven by time-resolved resonance fluorescence studies. We demonstrated near-background free single-photon emission with $g^{(2)}(0)=0.0191\pm0.0007$ and near-unity indistinguishability of 0.95$\pm$0.02. To best of our knowledge, it is the first demonstration of resonance fluorescence studies on the on-chip integrated ring resonators showing optical coherent control and on-demand indistinguishable single-photon generation. With the high degree of indistinguishability of the on-chip generated photons shown here, our structures could be used to realize various optical quantum-computing algorithms, interference of multiple photons, and the generation of photonic cluster states. We believe that our devices could be further improved in terms of performance by the better spatial alignment of the ring cavity mode and QD position to utilize larger Purcell enhancement and coupling efficiency while keeping high single-photon indistinguishability. Moreover, our structures could be straightforwardly monolithically integrated with other on-chip functionalities including beam-splitters, phase shifter, detectors, and other devices, suitable for handling large scale advanced quantum optics experiments on-chip. A potential to manufacture such circuits, combined with the high purity and potentially a high efficiency indistinguishable single-photon sources, opens a route toward fully integrated and thus scalable quantum information processing.  
	
	\begin{acknowledgement}
		The authors thank Silke Kuhn for fabricating the waveguide samples and also Hanna Salamon and Jakub Jasi\'nski for the initial characterization of the ring resonator devices. \L{}.D. acknowledges the financial support from the Alexander von Humboldt Foundation. S.-H. K. acknowledges the financial support from the National Research Foundation of Korea through the Korean Government Grant No. NRF-2019R1A2C4069587. We are furthermore grateful for the support by the State of Bavaria.
	\end{acknowledgement}
	
	\begin{suppinfo}
		The Supporting Information is available free of charge:\\
		Methods; Simulation details; Waveguide modes; Waveguide transmission losses; Circular Bragg reflector, inverse taper out-couplers and mode converters details; Ring design optimization; Extraction efficiency estimation; Two-photon interference histogram fitting and visibility correction; Performance of other ring resonator devices. 	
	\end{suppinfo}
	
	\bibliography{bib-Manuscript}

\providecommand{\latin}[1]{#1}
\makeatletter
\providecommand{\doi}
  {\begingroup\let\do\@makeother\dospecials
  \catcode`\{=1 \catcode`\}=2 \doi@aux}
\providecommand{\doi@aux}[1]{\endgroup\texttt{#1}}
\makeatother
\providecommand*\mcitethebibliography{\thebibliography}
\csname @ifundefined\endcsname{endmcitethebibliography}
  {\let\endmcitethebibliography\endthebibliography}{}
\begin{mcitethebibliography}{57}
\providecommand*\natexlab[1]{#1}
\providecommand*\mciteSetBstSublistMode[1]{}
\providecommand*\mciteSetBstMaxWidthForm[2]{}
\providecommand*\mciteBstWouldAddEndPuncttrue
  {\def\EndOfBibitem{\unskip.}}
\providecommand*\mciteBstWouldAddEndPunctfalse
  {\let\EndOfBibitem\relax}
\providecommand*\mciteSetBstMidEndSepPunct[3]{}
\providecommand*\mciteSetBstSublistLabelBeginEnd[3]{}
\providecommand*\EndOfBibitem{}
\mciteSetBstSublistMode{f}
\mciteSetBstMaxWidthForm{subitem}{(\alph{mcitesubitemcount})}
\mciteSetBstSublistLabelBeginEnd
  {\mcitemaxwidthsubitemform\space}
  {\relax}
  {\relax}

\bibitem[Santori \latin{et~al.}(2002)Santori, Fattal, Vuckovi{\'{c}}, Solomon,
  and Yamamoto]{Santori2002}
Santori,~C.; Fattal,~D.; Vuckovi{\'{c}},~J.; Solomon,~G.~S.; Yamamoto,~Y.
  {Indistinguishable photons from a single-photon device.} \emph{Nature}
  \textbf{2002}, \emph{419}, 594--7\relax
\mciteBstWouldAddEndPuncttrue
\mciteSetBstMidEndSepPunct{\mcitedefaultmidpunct}
{\mcitedefaultendpunct}{\mcitedefaultseppunct}\relax
\EndOfBibitem
\bibitem[Aharonovich \latin{et~al.}(2016)Aharonovich, Englund, and
  Toth]{Aharonovich2016}
Aharonovich,~I.; Englund,~D.; Toth,~M. {Solid-state single-photon emitters}.
  \emph{Nature Photonics} \textbf{2016}, \emph{10}, 631--641\relax
\mciteBstWouldAddEndPuncttrue
\mciteSetBstMidEndSepPunct{\mcitedefaultmidpunct}
{\mcitedefaultendpunct}{\mcitedefaultseppunct}\relax
\EndOfBibitem
\bibitem[Senellart \latin{et~al.}(2017)Senellart, Solomon, and
  White]{Senellart2017}
Senellart,~P.; Solomon,~G.; White,~A. {High-performance semiconductor
  quantum-dot single-photon sources}. \emph{Nature Nanotechnology}
  \textbf{2017}, \emph{12}, 1026--1039\relax
\mciteBstWouldAddEndPuncttrue
\mciteSetBstMidEndSepPunct{\mcitedefaultmidpunct}
{\mcitedefaultendpunct}{\mcitedefaultseppunct}\relax
\EndOfBibitem
\bibitem[Somaschi \latin{et~al.}(2016)Somaschi, Giesz, {De Santis}, Loredo,
  Almeida, Hornecker, Portalupi, Grange, Ant{\'{o}}n, Demory, G{\'{o}}mez,
  Sagnes, Lanzillotti-Kimura, Lema{\'{i}}tre, Auffeves, White, Lanco, and
  Senellart]{Somaschi2016}
Somaschi,~N. \latin{et~al.}  {Near-optimal single-photon sources in the solid
  state}. \emph{Nature Photonics} \textbf{2016}, \emph{10}, 340--345\relax
\mciteBstWouldAddEndPuncttrue
\mciteSetBstMidEndSepPunct{\mcitedefaultmidpunct}
{\mcitedefaultendpunct}{\mcitedefaultseppunct}\relax
\EndOfBibitem
\bibitem[Unsleber \latin{et~al.}(2016)Unsleber, He, Maier, Gerhardt, Lu, Pan,
  Kamp, Schneider, and H{\"{o}}fling]{Unsleber2016}
Unsleber,~S.; He,~Y.-M.; Maier,~S.; Gerhardt,~S.; Lu,~C.-Y.; Pan,~J.-W.;
  Kamp,~M.; Schneider,~C.; H{\"{o}}fling,~S. {Highly indistinguishable
  on-demand resonance fluorescence photons from a deterministic quantum dot
  micropillar device with 75{\%} extraction efficiency}. \emph{Optics express}
  \textbf{2016}, \emph{24}, 8539--8546\relax
\mciteBstWouldAddEndPuncttrue
\mciteSetBstMidEndSepPunct{\mcitedefaultmidpunct}
{\mcitedefaultendpunct}{\mcitedefaultseppunct}\relax
\EndOfBibitem
\bibitem[Wang \latin{et~al.}(2019)Wang, He, Chung, Hu, Yu, Chen, Ding, Chen,
  Qin, Yang, Liu, Duan, Li, Gerhardt, Winkler, Jurkat, Wang, Gregersen, Huo,
  Dai, Yu, H{\"{o}}fling, Lu, and Pan]{Wang2019}
Wang,~H. \latin{et~al.}  {Towards optimal single-photon sources from polarized
  microcavities}. \emph{Nature Photonics} \textbf{2019}, \emph{13},
  770--775\relax
\mciteBstWouldAddEndPuncttrue
\mciteSetBstMidEndSepPunct{\mcitedefaultmidpunct}
{\mcitedefaultendpunct}{\mcitedefaultseppunct}\relax
\EndOfBibitem
\bibitem[Pooley \latin{et~al.}(2012)Pooley, Ellis, Patel, Bennett, Chan,
  Farrer, Ritchie, and Shields]{Pooley2012}
Pooley,~M.~A.; Ellis,~D. J.~P.; Patel,~R.~B.; Bennett,~A.~J.; Chan,~K. H.~A.;
  Farrer,~I.; Ritchie,~D.~A.; Shields,~A.~J. {Controlled-NOT gate operating
  with single photons}. \emph{Applied Physics Letters} \textbf{2012},
  \emph{100}, 211103\relax
\mciteBstWouldAddEndPuncttrue
\mciteSetBstMidEndSepPunct{\mcitedefaultmidpunct}
{\mcitedefaultendpunct}{\mcitedefaultseppunct}\relax
\EndOfBibitem
\bibitem[Gazzano \latin{et~al.}(2013)Gazzano, Almeida, Nowak, Portalupi,
  Lema{\^{i}}tre, Sagnes, White, and Senellart]{Gazzano2013a}
Gazzano,~O.; Almeida,~M.~P.; Nowak,~A.~K.; Portalupi,~S.~L.;
  Lema{\^{i}}tre,~A.; Sagnes,~I.; White,~A.~G.; Senellart,~P. {Entangling
  Quantum-Logic Gate Operated with an Ultrabright Semiconductor Single-Photon
  Source}. \emph{Physical Review Letters} \textbf{2013}, \emph{110},
  250501\relax
\mciteBstWouldAddEndPuncttrue
\mciteSetBstMidEndSepPunct{\mcitedefaultmidpunct}
{\mcitedefaultendpunct}{\mcitedefaultseppunct}\relax
\EndOfBibitem
\bibitem[He \latin{et~al.}(2013)He, He, Wei, Wu, Atat{\"{u}}re, Schneider,
  H{\"{o}}fling, Kamp, Lu, and Pan]{He2013a}
He,~Y.-M.; He,~Y.; Wei,~Y.-J.; Wu,~D.; Atat{\"{u}}re,~M.; Schneider,~C.;
  H{\"{o}}fling,~S.; Kamp,~M.; Lu,~C.-Y.; Pan,~J.-W. {On-demand semiconductor
  single-photon source with near-unity indistinguishability}. \emph{Nature
  Nanotechnology} \textbf{2013}, \emph{8}, 213--217\relax
\mciteBstWouldAddEndPuncttrue
\mciteSetBstMidEndSepPunct{\mcitedefaultmidpunct}
{\mcitedefaultendpunct}{\mcitedefaultseppunct}\relax
\EndOfBibitem
\bibitem[Delteil \latin{et~al.}(2015)Delteil, Sun, Gao, Togan, Faelt, and
  Imamoğlu]{Delteil2015}
Delteil,~A.; Sun,~Z.; Gao,~W.-b.; Togan,~E.; Faelt,~S.; Imamoğlu,~A.
  {Generation of heralded entanglement between distant hole spins}.
  \emph{Nature Physics} \textbf{2015}, \emph{12}, 218--223\relax
\mciteBstWouldAddEndPuncttrue
\mciteSetBstMidEndSepPunct{\mcitedefaultmidpunct}
{\mcitedefaultendpunct}{\mcitedefaultseppunct}\relax
\EndOfBibitem
\bibitem[Stockill \latin{et~al.}(2017)Stockill, Stanley, Huthmacher, Clarke,
  Hugues, Miller, Matthiesen, {Le Gall}, and Atat{\"{u}}re]{Stockill2017}
Stockill,~R.; Stanley,~M.~J.; Huthmacher,~L.; Clarke,~E.; Hugues,~M.;
  Miller,~A.~J.; Matthiesen,~C.; {Le Gall},~C.; Atat{\"{u}}re,~M. {Phase-Tuned
  Entangled State Generation between Distant Spin Qubits}. \emph{Physical
  Review Letters} \textbf{2017}, \emph{119}, 010503\relax
\mciteBstWouldAddEndPuncttrue
\mciteSetBstMidEndSepPunct{\mcitedefaultmidpunct}
{\mcitedefaultendpunct}{\mcitedefaultseppunct}\relax
\EndOfBibitem
\bibitem[Nilsson \latin{et~al.}(2013)Nilsson, Stevenson, Chan, Skiba-Szymanska,
  Lucamarini, Ward, Bennett, Salter, Farrer, Ritchie, and Shields]{Nilsson2013}
Nilsson,~J.; Stevenson,~R.~M.; Chan,~K. H.~A.; Skiba-Szymanska,~J.;
  Lucamarini,~M.; Ward,~M.~B.; Bennett,~A.~J.; Salter,~C.~L.; Farrer,~I.;
  Ritchie,~D.~A.; Shields,~A.~J. {Quantum teleportation using a light-emitting
  diode}. \emph{Nature Photonics} \textbf{2013}, \emph{7}, 311--315\relax
\mciteBstWouldAddEndPuncttrue
\mciteSetBstMidEndSepPunct{\mcitedefaultmidpunct}
{\mcitedefaultendpunct}{\mcitedefaultseppunct}\relax
\EndOfBibitem
\bibitem[Reindl \latin{et~al.}(2018)Reindl, Huber, Schimpf, da~Silva, Rota,
  Huang, Zwiller, J{\"{o}}ns, Rastelli, and Trotta]{Reindl2018}
Reindl,~M.; Huber,~D.; Schimpf,~C.; da~Silva,~S. F.~C.; Rota,~M.~B.; Huang,~H.;
  Zwiller,~V.; J{\"{o}}ns,~K.~D.; Rastelli,~A.; Trotta,~R. {All-photonic
  quantum teleportation using on-demand solid-state quantum emitters}.
  \emph{Science Advances} \textbf{2018}, \emph{4}, eaau1255\relax
\mciteBstWouldAddEndPuncttrue
\mciteSetBstMidEndSepPunct{\mcitedefaultmidpunct}
{\mcitedefaultendpunct}{\mcitedefaultseppunct}\relax
\EndOfBibitem
\bibitem[Wang \latin{et~al.}(2019)Wang, Qin, Ding, Chen, Chen, You, He, Jiang,
  You, Wang, Schneider, Renema, H{\"{o}}fling, Lu, and Pan]{Wang2019-sampling}
Wang,~H.; Qin,~J.; Ding,~X.; Chen,~M.-C.; Chen,~S.; You,~X.; He,~Y.-M.;
  Jiang,~X.; You,~L.; Wang,~Z.; Schneider,~C.; Renema,~J.~J.;
  H{\"{o}}fling,~S.; Lu,~C.-Y.; Pan,~J.-W. {Boson Sampling with 20 Input
  Photons and a 60-Mode Interferometer in a 10$^{14}$ -Dimensional Hilbert
  Space}. \emph{Physical Review Letters} \textbf{2019}, \emph{123},
  250503\relax
\mciteBstWouldAddEndPuncttrue
\mciteSetBstMidEndSepPunct{\mcitedefaultmidpunct}
{\mcitedefaultendpunct}{\mcitedefaultseppunct}\relax
\EndOfBibitem
\bibitem[Flagg \latin{et~al.}(2009)Flagg, Muller, Robertson, Founta, Deppe,
  Xiao, Ma, Salamo, and Shih]{Flagg2009}
Flagg,~E.~B.; Muller,~A.; Robertson,~J.~W.; Founta,~S.; Deppe,~D.~G.; Xiao,~M.;
  Ma,~W.; Salamo,~G.~J.; Shih,~C.~K. {Resonantly driven coherent oscillations
  in a solid-state quantum emitter}. \emph{Nature Physics} \textbf{2009},
  \emph{5}, 203--207\relax
\mciteBstWouldAddEndPuncttrue
\mciteSetBstMidEndSepPunct{\mcitedefaultmidpunct}
{\mcitedefaultendpunct}{\mcitedefaultseppunct}\relax
\EndOfBibitem
\bibitem[Ates \latin{et~al.}(2009)Ates, Ulrich, Reitzenstein, L{\"{o}}ffler,
  Forchel, and Michler]{Ates2009}
Ates,~S.; Ulrich,~S.~M.; Reitzenstein,~S.; L{\"{o}}ffler,~A.; Forchel,~A.;
  Michler,~P. {Post-Selected Indistinguishable Photons from the Resonance
  Fluorescence of a Single Quantum Dot in a Microcavity}. \emph{Physical Review
  Letters} \textbf{2009}, \emph{103}, 167402\relax
\mciteBstWouldAddEndPuncttrue
\mciteSetBstMidEndSepPunct{\mcitedefaultmidpunct}
{\mcitedefaultendpunct}{\mcitedefaultseppunct}\relax
\EndOfBibitem
\bibitem[{Nick Vamivakas} \latin{et~al.}(2009){Nick Vamivakas}, Zhao, Lu, and
  Atat{\"{u}}re]{NickVamivakas2009}
{Nick Vamivakas},~A.; Zhao,~Y.; Lu,~C.-Y.; Atat{\"{u}}re,~M. {Spin-resolved
  quantum-dot resonance fluorescence}. \emph{Nature Physics} \textbf{2009},
  \emph{5}, 198--202\relax
\mciteBstWouldAddEndPuncttrue
\mciteSetBstMidEndSepPunct{\mcitedefaultmidpunct}
{\mcitedefaultendpunct}{\mcitedefaultseppunct}\relax
\EndOfBibitem
\bibitem[Iles-Smith \latin{et~al.}(2017)Iles-Smith, McCutcheon, Nazir, and
  M{\o}rk]{Iles-Smith2017}
Iles-Smith,~J.; McCutcheon,~D. P.~S.; Nazir,~A.; M{\o}rk,~J. {Phonon scattering
  inhibits simultaneous near-unity efficiency and indistinguishability in
  semiconductor single-photon sources}. \emph{Nature Photonics} \textbf{2017},
  \emph{11}, 521--526\relax
\mciteBstWouldAddEndPuncttrue
\mciteSetBstMidEndSepPunct{\mcitedefaultmidpunct}
{\mcitedefaultendpunct}{\mcitedefaultseppunct}\relax
\EndOfBibitem
\bibitem[Dietrich \latin{et~al.}(2016)Dietrich, Fiore, Thompson, Kamp, and
  H{\"{o}}fling]{Dietrich2016a}
Dietrich,~C.~P.; Fiore,~A.; Thompson,~M.~G.; Kamp,~M.; H{\"{o}}fling,~S. {GaAs
  integrated quantum photonics: Towards compact and multi-functional quantum
  photonic integrated circuits}. \emph{Laser {\&} Photonics Reviews}
  \textbf{2016}, \emph{10}, 870\relax
\mciteBstWouldAddEndPuncttrue
\mciteSetBstMidEndSepPunct{\mcitedefaultmidpunct}
{\mcitedefaultendpunct}{\mcitedefaultseppunct}\relax
\EndOfBibitem
\bibitem[Hepp \latin{et~al.}(2019)Hepp, Jetter, Portalupi, and
  Michler]{Hepp2019}
Hepp,~S.; Jetter,~M.; Portalupi,~S.~L.; Michler,~P. {Semiconductor Quantum Dots
  for Integrated Quantum Photonics}. \emph{Advanced Quantum Technologies}
  \textbf{2019}, \emph{1900020}, 1900020\relax
\mciteBstWouldAddEndPuncttrue
\mciteSetBstMidEndSepPunct{\mcitedefaultmidpunct}
{\mcitedefaultendpunct}{\mcitedefaultseppunct}\relax
\EndOfBibitem
\bibitem[J{\"{o}}ns \latin{et~al.}(2015)J{\"{o}}ns, Rengstl, Oster, Hargart,
  Heldmaier, Bounouar, Ulrich, Jetter, and Michler]{Jons2015}
J{\"{o}}ns,~K.~D.; Rengstl,~U.; Oster,~M.; Hargart,~F.; Heldmaier,~M.;
  Bounouar,~S.; Ulrich,~S.~M.; Jetter,~M.; Michler,~P. {Monolithic on-chip
  integration of semiconductor waveguides, beamsplitters and single-photon
  sources}. \emph{Journal of Physics D: Applied Physics} \textbf{2015},
  \emph{48}, 085101\relax
\mciteBstWouldAddEndPuncttrue
\mciteSetBstMidEndSepPunct{\mcitedefaultmidpunct}
{\mcitedefaultendpunct}{\mcitedefaultseppunct}\relax
\EndOfBibitem
\bibitem[Enderlin \latin{et~al.}(2012)Enderlin, Ota, Ohta, Kumagai, Ishida,
  Iwamoto, and Arakawa]{Enderlin2012}
Enderlin,~A.; Ota,~Y.; Ohta,~R.; Kumagai,~N.; Ishida,~S.; Iwamoto,~S.;
  Arakawa,~Y. {High guided mode–cavity mode coupling for an efficient
  extraction of spontaneous emission of a single quantum dot embedded in a
  photonic crystal nanobeam cavity}. \emph{Physical Review B} \textbf{2012},
  \emph{86}, 075314\relax
\mciteBstWouldAddEndPuncttrue
\mciteSetBstMidEndSepPunct{\mcitedefaultmidpunct}
{\mcitedefaultendpunct}{\mcitedefaultseppunct}\relax
\EndOfBibitem
\bibitem[Schwagmann \latin{et~al.}(2011)Schwagmann, Kalliakos, Farrer,
  Griffiths, Jones, Ritchie, and Shields]{Schwagmann2011}
Schwagmann,~A.; Kalliakos,~S.; Farrer,~I.; Griffiths,~J.~P.; Jones,~G. A.~C.;
  Ritchie,~D.~A.; Shields,~A.~J. {On-chip single photon emission from an
  integrated semiconductor quantum dot into a photonic crystal waveguide}.
  \emph{Applied Physics Letters} \textbf{2011}, \emph{99}, 261108\relax
\mciteBstWouldAddEndPuncttrue
\mciteSetBstMidEndSepPunct{\mcitedefaultmidpunct}
{\mcitedefaultendpunct}{\mcitedefaultseppunct}\relax
\EndOfBibitem
\bibitem[Arcari \latin{et~al.}(2014)Arcari, S{\"{o}}llner, Javadi, {Lindskov
  Hansen}, Mahmoodian, Liu, Thyrrestrup, Lee, Song, Stobbe, and
  Lodahl]{Arcari2014}
Arcari,~M.; S{\"{o}}llner,~I.; Javadi,~A.; {Lindskov Hansen},~S.;
  Mahmoodian,~S.; Liu,~J.; Thyrrestrup,~H.; Lee,~E.~H.; Song,~J.~D.;
  Stobbe,~S.; Lodahl,~P. {Near-Unity Coupling Efficiency of a Quantum Emitter
  to a Photonic Crystal Waveguide}. \emph{Physical Review Letters}
  \textbf{2014}, \emph{113}, 093603\relax
\mciteBstWouldAddEndPuncttrue
\mciteSetBstMidEndSepPunct{\mcitedefaultmidpunct}
{\mcitedefaultendpunct}{\mcitedefaultseppunct}\relax
\EndOfBibitem
\bibitem[Reithmaier \latin{et~al.}(2015)Reithmaier, Kaniber, Flassig,
  Lichtmannecker, M{\"{u}}ller, Andrejew, Vu{\v{c}}kovi{\'{c}}, Gross, and
  Finley]{Reithmaier2015}
Reithmaier,~G.; Kaniber,~M.; Flassig,~F.; Lichtmannecker,~S.; M{\"{u}}ller,~K.;
  Andrejew,~A.; Vu{\v{c}}kovi{\'{c}},~J.; Gross,~R.; Finley,~J.~J. {On-Chip
  Generation, Routing, and Detection of Resonance Fluorescence}. \emph{Nano
  Letters} \textbf{2015}, \emph{15}, 5208--5213\relax
\mciteBstWouldAddEndPuncttrue
\mciteSetBstMidEndSepPunct{\mcitedefaultmidpunct}
{\mcitedefaultendpunct}{\mcitedefaultseppunct}\relax
\EndOfBibitem
\bibitem[Davanco \latin{et~al.}(2017)Davanco, Liu, Sapienza, Zhang, {De Miranda
  Cardoso}, Verma, Mirin, Nam, Liu, and Srinivasan]{Davanco2017}
Davanco,~M.; Liu,~J.; Sapienza,~L.; Zhang,~C.-Z.; {De Miranda Cardoso},~J.~V.;
  Verma,~V.; Mirin,~R.; Nam,~S.~W.; Liu,~L.; Srinivasan,~K. {Heterogeneous
  integration for on-chip quantum photonic circuits with single quantum dot
  devices}. \emph{Nature Communications} \textbf{2017}, \emph{8}, 889\relax
\mciteBstWouldAddEndPuncttrue
\mciteSetBstMidEndSepPunct{\mcitedefaultmidpunct}
{\mcitedefaultendpunct}{\mcitedefaultseppunct}\relax
\EndOfBibitem
\bibitem[Elshaari \latin{et~al.}(2017)Elshaari, Zadeh, Fognini, Reimer, Dalacu,
  Poole, Zwiller, and J{\"{o}}ns]{Elshaari2017}
Elshaari,~A.~W.; Zadeh,~I.~E.; Fognini,~A.; Reimer,~M.~E.; Dalacu,~D.;
  Poole,~P.~J.; Zwiller,~V.; J{\"{o}}ns,~K.~D. {On-chip single photon filtering
  and multiplexing in hybrid quantum photonic circuits}. \emph{Nature
  Communications} \textbf{2017}, \emph{8}, 379\relax
\mciteBstWouldAddEndPuncttrue
\mciteSetBstMidEndSepPunct{\mcitedefaultmidpunct}
{\mcitedefaultendpunct}{\mcitedefaultseppunct}\relax
\EndOfBibitem
\bibitem[Kim \latin{et~al.}(2017)Kim, Aghaeimeibodi, Richardson, Leavitt,
  Englund, and Waks]{Kim2017}
Kim,~J.-H.; Aghaeimeibodi,~S.; Richardson,~C. J.~K.; Leavitt,~R.~P.;
  Englund,~D.; Waks,~E. {Hybrid Integration of Solid-State Quantum Emitters on
  a Silicon Photonic Chip}. \emph{Nano Letters} \textbf{2017}, \emph{17},
  7394\relax
\mciteBstWouldAddEndPuncttrue
\mciteSetBstMidEndSepPunct{\mcitedefaultmidpunct}
{\mcitedefaultendpunct}{\mcitedefaultseppunct}\relax
\EndOfBibitem
\bibitem[Ellis \latin{et~al.}(2018)Ellis, Bennett, Dangel, Lee, Griffiths,
  Mitchell, Paraiso, Spencer, Ritchie, and Shields]{Ellis2018}
Ellis,~D. J.~P.; Bennett,~A.~J.; Dangel,~C.; Lee,~J.~P.; Griffiths,~J.~P.;
  Mitchell,~T.~A.; Paraiso,~T.-K.; Spencer,~P.; Ritchie,~D.~A.; Shields,~A.~J.
  {Independent indistinguishable quantum light sources on a reconfigurable
  photonic integrated circuit}. \emph{Applied Physics Letters} \textbf{2018},
  \emph{112}, 211104\relax
\mciteBstWouldAddEndPuncttrue
\mciteSetBstMidEndSepPunct{\mcitedefaultmidpunct}
{\mcitedefaultendpunct}{\mcitedefaultseppunct}\relax
\EndOfBibitem
\bibitem[Midolo \latin{et~al.}(2017)Midolo, Hansen, Zhang, Papon, Schott,
  Ludwig, Wieck, Lodahl, and Stobbe]{Midolo2017}
Midolo,~L.; Hansen,~S.~L.; Zhang,~W.; Papon,~C.; Schott,~R.; Ludwig,~A.;
  Wieck,~A.~D.; Lodahl,~P.; Stobbe,~S. Electro-optic routing of photons from a
  single quantum dot in photonic integrated circuits. \emph{Opt. Express}
  \textbf{2017}, \emph{25}, 33514--33526\relax
\mciteBstWouldAddEndPuncttrue
\mciteSetBstMidEndSepPunct{\mcitedefaultmidpunct}
{\mcitedefaultendpunct}{\mcitedefaultseppunct}\relax
\EndOfBibitem
\bibitem[Wang \latin{et~al.}(2014)Wang, Santamato, Jiang, Bonneau, Engin,
  Silverstone, Lermer, Beetz, Kamp, H{\"{o}}fling, Tanner, Natarajan, Hadfield,
  Dorenbos, Zwiller, O'Brien, and Thompson]{Wang2014a}
Wang,~J. \latin{et~al.}  {Gallium arsenide (GaAs) quantum photonic waveguide
  circuits}. \emph{Optics Communications} \textbf{2014}, \emph{327},
  49--55\relax
\mciteBstWouldAddEndPuncttrue
\mciteSetBstMidEndSepPunct{\mcitedefaultmidpunct}
{\mcitedefaultendpunct}{\mcitedefaultseppunct}\relax
\EndOfBibitem
\bibitem[Prtljaga \latin{et~al.}(2014)Prtljaga, Coles, O'Hara, Royall, Clarke,
  Fox, and Skolnick]{Prtljaga2014}
Prtljaga,~N.; Coles,~R.~J.; O'Hara,~J.; Royall,~B.; Clarke,~E.; Fox,~A.~M.;
  Skolnick,~M.~S. {Monolithic integration of a quantum emitter with a compact
  on-chip beam-splitter}. \emph{Applied Physics Letters} \textbf{2014},
  \emph{104}, 231107\relax
\mciteBstWouldAddEndPuncttrue
\mciteSetBstMidEndSepPunct{\mcitedefaultmidpunct}
{\mcitedefaultendpunct}{\mcitedefaultseppunct}\relax
\EndOfBibitem
\bibitem[Elshaari \latin{et~al.}(2018)Elshaari, B{\"{u}}y{\"{u}}k{\"{o}}zer,
  Zadeh, Lettner, Zhao, Sch{\"{o}}ll, Gyger, Reimer, Dalacu, Poole, J{\"{o}}ns,
  and Zwiller]{Elshaari2018}
Elshaari,~A.~W.; B{\"{u}}y{\"{u}}k{\"{o}}zer,~E.; Zadeh,~I.~E.; Lettner,~T.;
  Zhao,~P.; Sch{\"{o}}ll,~E.; Gyger,~S.; Reimer,~M.~E.; Dalacu,~D.;
  Poole,~P.~J.; J{\"{o}}ns,~K.~D.; Zwiller,~V. {Strain-Tunable Quantum
  Integrated Photonics}. \emph{Nano Letters} \textbf{2018}, \emph{18},
  7969--7976\relax
\mciteBstWouldAddEndPuncttrue
\mciteSetBstMidEndSepPunct{\mcitedefaultmidpunct}
{\mcitedefaultendpunct}{\mcitedefaultseppunct}\relax
\EndOfBibitem
\bibitem[Aghaeimeibodi \latin{et~al.}(2019)Aghaeimeibodi, Kim, Lee, Buyukkaya,
  Richardson, and Waks]{Aghaeimeibodi2019}
Aghaeimeibodi,~S.; Kim,~J.-H.; Lee,~C.-M.; Buyukkaya,~M.~A.; Richardson,~C.;
  Waks,~E. {Silicon photonic add-drop filter for quantum emitters}.
  \emph{Optics Express} \textbf{2019}, \emph{27}, 16882\relax
\mciteBstWouldAddEndPuncttrue
\mciteSetBstMidEndSepPunct{\mcitedefaultmidpunct}
{\mcitedefaultendpunct}{\mcitedefaultseppunct}\relax
\EndOfBibitem
\bibitem[Kaniber \latin{et~al.}(2016)Kaniber, Flassig, Reithmaier, Gross, and
  Finley]{Kaniber2016}
Kaniber,~M.; Flassig,~F.; Reithmaier,~G.; Gross,~R.; Finley,~J.~J. {Integrated
  superconducting detectors on semiconductors for quantum optics applications}.
  \emph{Applied Physics B} \textbf{2016}, \emph{122}, 115\relax
\mciteBstWouldAddEndPuncttrue
\mciteSetBstMidEndSepPunct{\mcitedefaultmidpunct}
{\mcitedefaultendpunct}{\mcitedefaultseppunct}\relax
\EndOfBibitem
\bibitem[Makhonin \latin{et~al.}(2014)Makhonin, Dixon, Coles, Royall, Luxmoore,
  Clarke, Hugues, Skolnick, and Fox]{Makhonin2014}
Makhonin,~M.~N.; Dixon,~J.~E.; Coles,~R.~J.; Royall,~B.; Luxmoore,~I.~J.;
  Clarke,~E.; Hugues,~M.; Skolnick,~M.~S.; Fox,~A.~M. {Waveguide Coupled
  Resonance Fluorescence from On-Chip Quantum Emitter}. \emph{Nano Letters}
  \textbf{2014}, \emph{14}, 6997--7002\relax
\mciteBstWouldAddEndPuncttrue
\mciteSetBstMidEndSepPunct{\mcitedefaultmidpunct}
{\mcitedefaultendpunct}{\mcitedefaultseppunct}\relax
\EndOfBibitem
\bibitem[Kalliakos \latin{et~al.}(2016)Kalliakos, Brody, Bennett, Ellis,
  Skiba-Szymanska, Farrer, Griffiths, Ritchie, and Shields]{Kalliakos2016}
Kalliakos,~S.; Brody,~Y.; Bennett,~A.~J.; Ellis,~D. J.~P.; Skiba-Szymanska,~J.;
  Farrer,~I.; Griffiths,~J.~P.; Ritchie,~D.~A.; Shields,~A.~J. {Enhanced
  indistinguishability of in-plane single photons by resonance fluorescence on
  an integrated quantum dot}. \emph{Applied Physics Letters} \textbf{2016},
  \emph{109}, 151112\relax
\mciteBstWouldAddEndPuncttrue
\mciteSetBstMidEndSepPunct{\mcitedefaultmidpunct}
{\mcitedefaultendpunct}{\mcitedefaultseppunct}\relax
\EndOfBibitem
\bibitem[Schwartz \latin{et~al.}(2016)Schwartz, Rengstl, Herzog, Paul, Kettler,
  Portalupi, Jetter, and Michler]{Schwartz2016}
Schwartz,~M.; Rengstl,~U.; Herzog,~T.; Paul,~M.; Kettler,~J.; Portalupi,~S.~L.;
  Jetter,~M.; Michler,~P. {Generation, guiding and splitting of triggered
  single photons from a resonantly excited quantum dot in a photonic circuit}.
  \emph{Optics Express} \textbf{2016}, \emph{24}, 3089--3094\relax
\mciteBstWouldAddEndPuncttrue
\mciteSetBstMidEndSepPunct{\mcitedefaultmidpunct}
{\mcitedefaultendpunct}{\mcitedefaultseppunct}\relax
\EndOfBibitem
\bibitem[Kir{\v{s}}anskė \latin{et~al.}(2017)Kir{\v{s}}anskė, Thyrrestrup,
  Daveau, Dree{\ss}en, Pregnolato, Midolo, Tighineanu, Javadi, Stobbe, Schott,
  Ludwig, Wieck, Park, Song, Kuhlmann, S{\"{o}}llner, L{\"{o}}bl, Warburton,
  and Lodahl]{Kirsanske2017b}
Kir{\v{s}}anskė,~G. \latin{et~al.}  {Indistinguishable and efficient single
  photons from a quantum dot in a planar nanobeam waveguide}. \emph{Physical
  Review B} \textbf{2017}, \emph{96}, 165306\relax
\mciteBstWouldAddEndPuncttrue
\mciteSetBstMidEndSepPunct{\mcitedefaultmidpunct}
{\mcitedefaultendpunct}{\mcitedefaultseppunct}\relax
\EndOfBibitem
\bibitem[Liu \latin{et~al.}(2018)Liu, Brash, O'Hara, Martins, Phillips, Coles,
  Royall, Clarke, Bentham, Prtljaga, Itskevich, Wilson, Skolnick, and
  Fox]{Liu2018}
Liu,~F.; Brash,~A.~J.; O'Hara,~J.; Martins,~L. M. P.~P.; Phillips,~C.~L.;
  Coles,~R.~J.; Royall,~B.; Clarke,~E.; Bentham,~C.; Prtljaga,~N.;
  Itskevich,~I.~E.; Wilson,~L.~R.; Skolnick,~M.~S.; Fox,~A.~M. {High Purcell
  factor generation of indistinguishable on-chip single photons}. \emph{Nature
  Nanotechnology} \textbf{2018}, \emph{13}, 835--840\relax
\mciteBstWouldAddEndPuncttrue
\mciteSetBstMidEndSepPunct{\mcitedefaultmidpunct}
{\mcitedefaultendpunct}{\mcitedefaultseppunct}\relax
\EndOfBibitem
\bibitem[Dusanowski \latin{et~al.}(2019)Dusanowski, Kwon, Schneider, and
  H{\"{o}}fling]{Dusanowski2019}
Dusanowski,~{\L}.; Kwon,~S.-h.; Schneider,~C.; H{\"{o}}fling,~S. {Near-Unity
  Indistinguishability Single Photon Source for Large-Scale Integrated Quantum
  Optics}. \emph{Physical Review Letters} \textbf{2019}, \emph{122},
  173602\relax
\mciteBstWouldAddEndPuncttrue
\mciteSetBstMidEndSepPunct{\mcitedefaultmidpunct}
{\mcitedefaultendpunct}{\mcitedefaultseppunct}\relax
\EndOfBibitem
\bibitem[Lund-Hansen \latin{et~al.}(2008)Lund-Hansen, Stobbe, Julsgaard,
  Thyrrestrup, S{\"{u}}nner, Kamp, Forchel, and Lodahl]{Lund-Hansen2008}
Lund-Hansen,~T.; Stobbe,~S.; Julsgaard,~B.; Thyrrestrup,~H.; S{\"{u}}nner,~T.;
  Kamp,~M.; Forchel,~A.; Lodahl,~P. {Experimental Realization of Highly
  Efficient Broadband Coupling of Single Quantum Dots to a Photonic Crystal
  Waveguide}. \emph{Physical Review Letters} \textbf{2008}, \emph{101},
  113903\relax
\mciteBstWouldAddEndPuncttrue
\mciteSetBstMidEndSepPunct{\mcitedefaultmidpunct}
{\mcitedefaultendpunct}{\mcitedefaultseppunct}\relax
\EndOfBibitem
\bibitem[Stepanov \latin{et~al.}(2015)Stepanov, Delga, Zang, Bleuse, Dupuy,
  Peinke, Lalanne, G{\'{e}}rard, and Claudon]{Stepanov2015}
Stepanov,~P.; Delga,~A.; Zang,~X.; Bleuse,~J.; Dupuy,~E.; Peinke,~E.;
  Lalanne,~P.; G{\'{e}}rard,~J.-M.; Claudon,~J. {Quantum dot spontaneous
  emission control in a ridge waveguide}. \emph{Applied Physics Letters}
  \textbf{2015}, \emph{106}, 041112\relax
\mciteBstWouldAddEndPuncttrue
\mciteSetBstMidEndSepPunct{\mcitedefaultmidpunct}
{\mcitedefaultendpunct}{\mcitedefaultseppunct}\relax
\EndOfBibitem
\bibitem[Fattah~poor \latin{et~al.}(2013)Fattah~poor, Hoang, Midolo, Dietrich,
  Li, Linfield, Schouwenberg, Xia, Pagliano, van Otten, and
  Fiore]{Fattahpoor2013}
Fattah~poor,~S.; Hoang,~T.~B.; Midolo,~L.; Dietrich,~C.~P.; Li,~L.~H.;
  Linfield,~E.~H.; Schouwenberg,~J. F.~P.; Xia,~T.; Pagliano,~F.~M.; van
  Otten,~F. W.~M.; Fiore,~A. {Efficient coupling of single photons to
  ridge-waveguide photonic integrated circuits}. \emph{Applied Physics Letters}
  \textbf{2013}, \emph{102}, 131105\relax
\mciteBstWouldAddEndPuncttrue
\mciteSetBstMidEndSepPunct{\mcitedefaultmidpunct}
{\mcitedefaultendpunct}{\mcitedefaultseppunct}\relax
\EndOfBibitem
\bibitem[Schwartz \latin{et~al.}(2018)Schwartz, Schmidt, Rengstl, Hornung,
  Hepp, Portalupi, Llin, Jetter, Siegel, and Michler]{Schwartz2018}
Schwartz,~M.; Schmidt,~E.; Rengstl,~U.; Hornung,~F.; Hepp,~S.;
  Portalupi,~S.~L.; Llin,~K.; Jetter,~M.; Siegel,~M.; Michler,~P. {Fully
  On-Chip Single-Photon Hanbury-Brown and Twiss Experiment on a Monolithic
  Semiconductor–Superconductor Platform}. \emph{Nano Letters} \textbf{2018},
  \emph{18}, 6892--6897\relax
\mciteBstWouldAddEndPuncttrue
\mciteSetBstMidEndSepPunct{\mcitedefaultmidpunct}
{\mcitedefaultendpunct}{\mcitedefaultseppunct}\relax
\EndOfBibitem
\bibitem[Reigue \latin{et~al.}(2017)Reigue, Iles-Smith, Lux, Monniello,
  Bernard, Margaillan, Lemaitre, Martinez, McCutcheon, M{\o}rk, Hostein, and
  Voliotis]{Reigue2017}
Reigue,~A.; Iles-Smith,~J.; Lux,~F.; Monniello,~L.; Bernard,~M.;
  Margaillan,~F.; Lemaitre,~A.; Martinez,~A.; McCutcheon,~D.~P.; M{\o}rk,~J.;
  Hostein,~R.; Voliotis,~V. {Probing Electron-Phonon Interaction through
  Two-Photon Interference in Resonantly Driven Semiconductor Quantum Dots}.
  \emph{Physical Review Letters} \textbf{2017}, \emph{118}, 233602\relax
\mciteBstWouldAddEndPuncttrue
\mciteSetBstMidEndSepPunct{\mcitedefaultmidpunct}
{\mcitedefaultendpunct}{\mcitedefaultseppunct}\relax
\EndOfBibitem
\bibitem[Hepp \latin{et~al.}(2018)Hepp, Bauer, Hornung, Schwartz, Portalupi,
  Jetter, and Michler]{Hepp2018}
Hepp,~S.; Bauer,~S.; Hornung,~F.; Schwartz,~M.; Portalupi,~S.~L.; Jetter,~M.;
  Michler,~P. {Bragg grating cavities embedded into nano-photonic waveguides
  for Purcell enhanced quantum dot emission}. \emph{Optics Express}
  \textbf{2018}, \emph{26}, 30614\relax
\mciteBstWouldAddEndPuncttrue
\mciteSetBstMidEndSepPunct{\mcitedefaultmidpunct}
{\mcitedefaultendpunct}{\mcitedefaultseppunct}\relax
\EndOfBibitem
\bibitem[Rab(2007)]{Rabus2007}
\emph{Integrated Ring Resonators: The Compendium}; Springer Berlin Heidelberg:
  Berlin, Heidelberg, 2007; pp 3--40\relax
\mciteBstWouldAddEndPuncttrue
\mciteSetBstMidEndSepPunct{\mcitedefaultmidpunct}
{\mcitedefaultendpunct}{\mcitedefaultseppunct}\relax
\EndOfBibitem
\bibitem[Wen \latin{et~al.}(2011)Wen, Kuzucu, Hou, Lipson, and Gaeta]{Wen2011}
Wen,~Y.~H.; Kuzucu,~O.; Hou,~T.; Lipson,~M.; Gaeta,~A.~L. {All-optical
  switching of a single resonance in silicon ring resonators}. \emph{Optics
  Letters} \textbf{2011}, \emph{36}, 1413\relax
\mciteBstWouldAddEndPuncttrue
\mciteSetBstMidEndSepPunct{\mcitedefaultmidpunct}
{\mcitedefaultendpunct}{\mcitedefaultseppunct}\relax
\EndOfBibitem
\bibitem[Engin \latin{et~al.}(2013)Engin, Bonneau, Natarajan, Clark, Tanner,
  Hadfield, Dorenbos, Zwiller, Ohira, Suzuki, Yoshida, Iizuka, Ezaki, O'Brien,
  and Thompson]{Engin2013}
Engin,~E.; Bonneau,~D.; Natarajan,~C.~M.; Clark,~A.~S.; Tanner,~M.~G.;
  Hadfield,~R.~H.; Dorenbos,~S.~N.; Zwiller,~V.; Ohira,~K.; Suzuki,~N.;
  Yoshida,~H.; Iizuka,~N.; Ezaki,~M.; O'Brien,~J.~L.; Thompson,~M.~G. {Photon
  pair generation in a silicon micro-ring resonator with reverse bias
  enhancement}. \emph{Optics Express} \textbf{2013}, \emph{21}, 27826\relax
\mciteBstWouldAddEndPuncttrue
\mciteSetBstMidEndSepPunct{\mcitedefaultmidpunct}
{\mcitedefaultendpunct}{\mcitedefaultseppunct}\relax
\EndOfBibitem
\bibitem[Faraon \latin{et~al.}(2011)Faraon, Barclay, Santori, Fu, and
  Beausoleil]{Faraon2011}
Faraon,~A.; Barclay,~P.~E.; Santori,~C.; Fu,~K.-M.~C.; Beausoleil,~R.~G.
  {Resonant enhancement of the zero-phonon emission from a colour centre in a
  diamond cavity}. \emph{Nature Photonics} \textbf{2011}, \emph{5},
  301--305\relax
\mciteBstWouldAddEndPuncttrue
\mciteSetBstMidEndSepPunct{\mcitedefaultmidpunct}
{\mcitedefaultendpunct}{\mcitedefaultseppunct}\relax
\EndOfBibitem
\bibitem[{Kuan Pei Yap} \latin{et~al.}(2009){Kuan Pei Yap}, Delage, Lapointe,
  Lamontagne, Schmid, Waldron, Syrett, and Janz]{KuanPeiYap2009}
{Kuan Pei Yap},; Delage,~A.; Lapointe,~J.; Lamontagne,~B.; Schmid,~J.;
  Waldron,~P.; Syrett,~B.; Janz,~S. {Correlation of Scattering Loss, Sidewall
  Roughness and Waveguide Width in Silicon-on-Insulator (SOI) Ridge
  Waveguides}. \emph{Journal of Lightwave Technology} \textbf{2009}, \emph{27},
  3999--4008\relax
\mciteBstWouldAddEndPuncttrue
\mciteSetBstMidEndSepPunct{\mcitedefaultmidpunct}
{\mcitedefaultendpunct}{\mcitedefaultseppunct}\relax
\EndOfBibitem
\bibitem[Gerhardt \latin{et~al.}(2018)Gerhardt, Iles-Smith, McCutcheon, He,
  Unsleber, Betzold, Gregersen, M{\o}rk, H{\"{o}}fling, and
  Schneider]{Gerhardt2018}
Gerhardt,~S.; Iles-Smith,~J.; McCutcheon,~D. P.~S.; He,~Y.-M.; Unsleber,~S.;
  Betzold,~S.; Gregersen,~N.; M{\o}rk,~J.; H{\"{o}}fling,~S.; Schneider,~C.
  {Intrinsic and environmental effects on the interference properties of a
  high-performance quantum dot single-photon source}. \emph{Physical Review B}
  \textbf{2018}, \emph{97}, 195432\relax
\mciteBstWouldAddEndPuncttrue
\mciteSetBstMidEndSepPunct{\mcitedefaultmidpunct}
{\mcitedefaultendpunct}{\mcitedefaultseppunct}\relax
\EndOfBibitem
\bibitem[Rengstl \latin{et~al.}(2015)Rengstl, Schwartz, Herzog, Hargart, Paul,
  Portalupi, Jetter, and Michler]{Rengstl2015}
Rengstl,~U.; Schwartz,~M.; Herzog,~T.; Hargart,~F.; Paul,~M.; Portalupi,~S.~L.;
  Jetter,~M.; Michler,~P. {On-chip beamsplitter operation on single photons
  from quasi-resonantly excited quantum dots embedded in GaAs rib waveguides}.
  \emph{Applied Physics Letters} \textbf{2015}, \emph{107}, 021101\relax
\mciteBstWouldAddEndPuncttrue
\mciteSetBstMidEndSepPunct{\mcitedefaultmidpunct}
{\mcitedefaultendpunct}{\mcitedefaultseppunct}\relax
\EndOfBibitem
\bibitem[Varnava \latin{et~al.}(2008)Varnava, Browne, and Rudolph]{Varnava2008}
Varnava,~M.; Browne,~D.~E.; Rudolph,~T. {How Good Must Single Photon Sources
  and Detectors Be for Efficient Linear Optical Quantum Computation?}
  \emph{Physical Review Letters} \textbf{2008}, \emph{100}, 060502\relax
\mciteBstWouldAddEndPuncttrue
\mciteSetBstMidEndSepPunct{\mcitedefaultmidpunct}
{\mcitedefaultendpunct}{\mcitedefaultseppunct}\relax
\EndOfBibitem
\bibitem[Fowler \latin{et~al.}(2009)Fowler, Stephens, and
  Groszkowski]{Fowler2009}
Fowler,~A.~G.; Stephens,~A.~M.; Groszkowski,~P. {High-threshold universal
  quantum computation on the surface code}. \emph{Physical Review A}
  \textbf{2009}, \emph{80}, 052312\relax
\mciteBstWouldAddEndPuncttrue
\mciteSetBstMidEndSepPunct{\mcitedefaultmidpunct}
{\mcitedefaultendpunct}{\mcitedefaultseppunct}\relax
\EndOfBibitem
\end{mcitethebibliography}


\providecommand{\latin}[1]{#1}
\makeatletter
\providecommand{\doi}
  {\begingroup\let\do\@makeother\dospecials
  \catcode`\{=1 \catcode`\}=2 \doi@aux}
\providecommand{\doi@aux}[1]{\endgroup\texttt{#1}}
\makeatother
\providecommand*\mcitethebibliography{\thebibliography}
\csname @ifundefined\endcsname{endmcitethebibliography}
  {\let\endmcitethebibliography\endthebibliography}{}
\begin{mcitethebibliography}{0}
\providecommand*\natexlab[1]{#1}
\providecommand*\mciteSetBstSublistMode[1]{}
\providecommand*\mciteSetBstMaxWidthForm[2]{}
\providecommand*\mciteBstWouldAddEndPuncttrue
  {\def\EndOfBibitem{\unskip.}}
\providecommand*\mciteBstWouldAddEndPunctfalse
  {\let\EndOfBibitem\relax}
\providecommand*\mciteSetBstMidEndSepPunct[3]{}
\providecommand*\mciteSetBstSublistLabelBeginEnd[3]{}
\providecommand*\EndOfBibitem{}
\mciteSetBstSublistMode{f}
\mciteSetBstMaxWidthForm{subitem}{(\alph{mcitesubitemcount})}
\mciteSetBstSublistLabelBeginEnd
  {\mcitemaxwidthsubitemform\space}
  {\relax}
  {\relax}

\end{mcitethebibliography}
	
	\newpage
	\section*{Graphic for Table of Content only}
	
	\begin{figure}
		\includegraphics[width=4.5in]{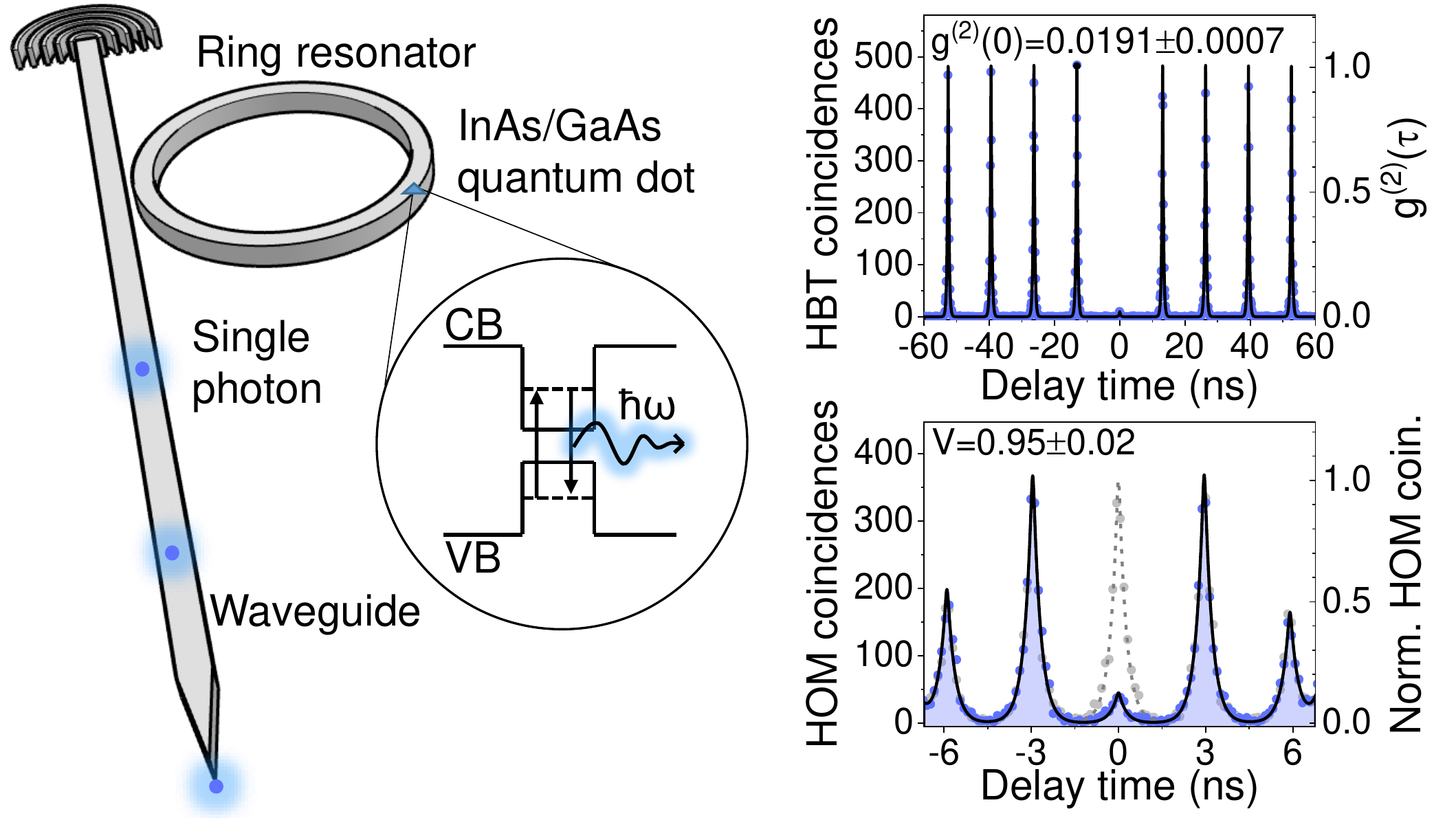}
	\end{figure}
	
\end{document}


\maketitle
	
	\renewcommand\thefigure{S\arabic{figure}} 
	\newpage

	\section{Methods}
\begin{figure}[!h]
		\includegraphics[width=6.5in]{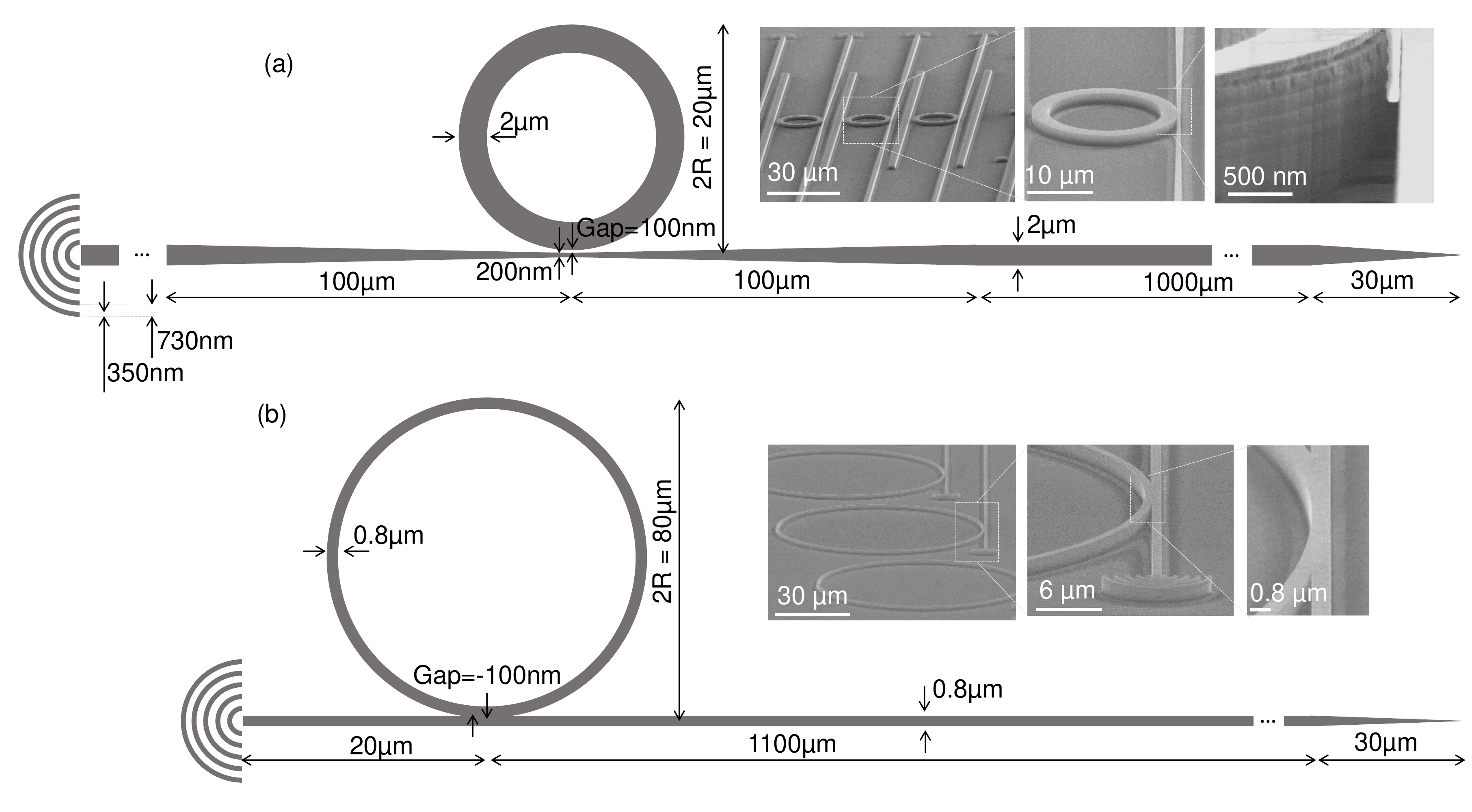}
		\caption{\label{fig:struc} Layout scheme of the investigated rings with (a)~10~$\mu$m radius and (b)~40~$\mu$m radius. Insets: Scanning electron microscope images of the fabricated devices. In the case of 10~$\mu$m radius ring nominal 100~nm gap is not perfectly etched through the GaAs cavity layer.}
	\end{figure}

	\textbf{Sample structure.} To fabricate our integrated single-photon source waveguide device we use a semiconductor sample which contains self-assembled In(Ga)As QDs grew by the Stranski-Krastanow method at the centre of a planar GaAs microcavity. The lower and upper cavity mirrors contain 24 and 5 pairs of Al$_{0.9}$Ga$_{0.1}$As/GaAs $\lambda$/4-layers, respectively, yielding a vertical confinement quality factor of $\sim$200. To fabricate ridge waveguides, the top mirror layer along with GaAs cavity is etched down, forming the ridges with a height of $\sim$1.3~$\mu$m. Ridges have been defined by e-beam lithography and dry etching. After processing, the sample was cleaved perpendicularly to the WGs to get clear side access to the ridge tapered-out-coupler facets. In Fig.~\ref{fig:struc}(a) and (b) layout schemes and SEM images of the fabricated ring resonators are shown.
	
	\textbf{Optical set-up.} For all experiments, the sample is kept in a high-stability closed cryostat at low temperatures of around $T=$~4.2~K. The cryostat is equipped with two optical windows allowing for access from both side and top of the sample via two independent perpendicularly aligned optical paths. QDs embedded into WG rings are excited from the top trough a microscope objective with numerical aperture NA~=~0.4, while the emission signal is detected from the side of the WG with a second objective with the same NA. For the resonance fluorescence experiments, in the side detection path, a combination of a linear polarizer aligned with TE WG mode and the spatial filter is used based on two 50~cm focal length lenses and 2D variable size slit. The collected light is then analyzed by a high-resolution monochromator equipped with a liquid nitrogen-cooled low-noise charge-coupled device detector, featuring a spectral resolution of $\sim$30~$\mu$eV. For non-resonant PL experiments, a 660~nm continuous-wave laser is used while for resonance fluorescence investigations a tunable mode-locked Ti:Sapphire laser with a repetition rate of 76~MHz and pulse width of around 2~ps is used. 
	
	\textbf{Purcell factor evaluation.}
	For a quantitative analysis of the Purcell enhancement, we must consider how the emitter is coupled to the vacuum field. The total spontaneous emission rate of an emitter coupled to a resonator, relative to the emitter in bulk, is enhanced by factor~\cite{Gayral2008,Munsch2009} $F_p+\Gamma_{leak}/\Gamma_{0}$, where $F_p$ is Purcell factor, $\Gamma_{leak}$ is emission rate into the leaky modes (outside the cavity mode), $\Gamma_{0}$ is emission rate in bulk. Following the above, for trustworthy Purcell factor evaluation, an exact knowledge of the $\Gamma_{leak}/\Gamma_{0}$ for a given photonic system is required. In structures where emission rate into the leaky modes is known to be equal to the emission rate in the bulk, such as micro-pillar cavities, $\Gamma_{leak}/\Gamma_{0}$ equal to 1 is usually assumed~\cite{Gayral2008,Munsch2009}. For photonic crystal cavities, however,  $\Gamma_{leak}/\Gamma_{0}$ can be significantly smaller than 1~\cite{Fujita2005}. In the ring resonator devices based on DBR waveguides, $\Gamma_{leak}/\Gamma_{0}$ is not ambiguously known. Given, that $0<\Gamma_{leak}/\Gamma_{0}<1$ Purcell factor in the range of 1.4-2.4 (1.7-2.7) can be extracted for device no.1 (no.2). The decay time $\tau$ data in Figure 3(b) is fit using formula~\cite{Gayral2008}
	\begin{equation*}
	\tau(\Delta E) = \frac{1/\Gamma_{0}}{\Gamma_{leak} / \Gamma_{0}+F_p \left[ 1+4 \left(\frac{Q\Delta E}{E_c}\right)^2 \right]^{-1}},
	\end{equation*} 
	assuming that the emission rate into the cavity mode follows a Lorentzian dependence with respect to the QD-cavity energy detuning $\Delta E$, where $Q$ is cavity quality factor, $E_c$ is cavity energy.
	
	\textbf{Auto-correlation and two-photon interference experiment details.}
	To characterize investigated source purity and indistinguishability, the resonance fluorescence signal is passed through a monochromator to filter out (spectral width $\sim$50~pm/70~$\mu$eV) a broader laser profile and phonon sidebands, and then coupled into a single-mode polarization-maintaining fiber. Next, light is introduced into a fiber-based unbalanced Mach-Zehnder interferometer for the two-photon interference measurements in the Hong-Ou-Mandel (HOM) configuration. For the auto-correlation measurements, one of the interferometer arms is blocked. Outputs ports are coupled to the pair of single-photon counting avalanche detectors (APD) with a 350~ps temporal resolution or superconducting detectors (SSPD) with a 30~ps resolution. The photon correlation events are acquired by a multi-channel picosecond event timer. For the time-resolved experiments, a fast SSPD is used.
	
	\newpage
	\section{Simulations details} 
	\label{app:1} 
	We calculated a coupling efficiency and Purcell factor by three-dimensional finite-difference-time-domain (FDTD) method. We used a combination of home-made FDTD and Lumerical software. The QD emitter is modelled by a linearly polarized dipole source aligned along the transverse-electrically (TE) polarized waveguide mode. The dipole is usually placed at the maximum of the fundamental TE bent ring waveguide mode. The refractive indexes of GaAs and AlGaAs are set to 3.5652 and 3.0404, respectively. To represent infinite free space in the simulations, the uniaxial perfectly matched layer (UPML) was used as the absorbing boundary condition. On the other hand, to calculate the electric field distribution of the waveguide modes, a periodic boundary condition is used for the waveguide direction instead of the UPML.
	
	The overall bus WG coupling efficiency is defined as the probability to detected one photon emitted from a QD placed in the ring resonator in the fundamental mode of the bus WG. The total QD-bus-WG coupling efficiency consists of two contributions: (i)~QD to ring mode coupling and (ii)~ring-mode to bus waveguide coupling. The QD-ring coupling efficiency is defined as the ratio of emission coupled into the ring resonator mode and the total emission from a QD. The ring-bus-WG efficiency is defined as the ratio between the energy coupled into the bus WG and the total energy in the ring cavity mode.
	
	\newpage
	\section{Waveguide modes}
\begin{figure}[h]
		\centering
		\includegraphics[width=6.5in]{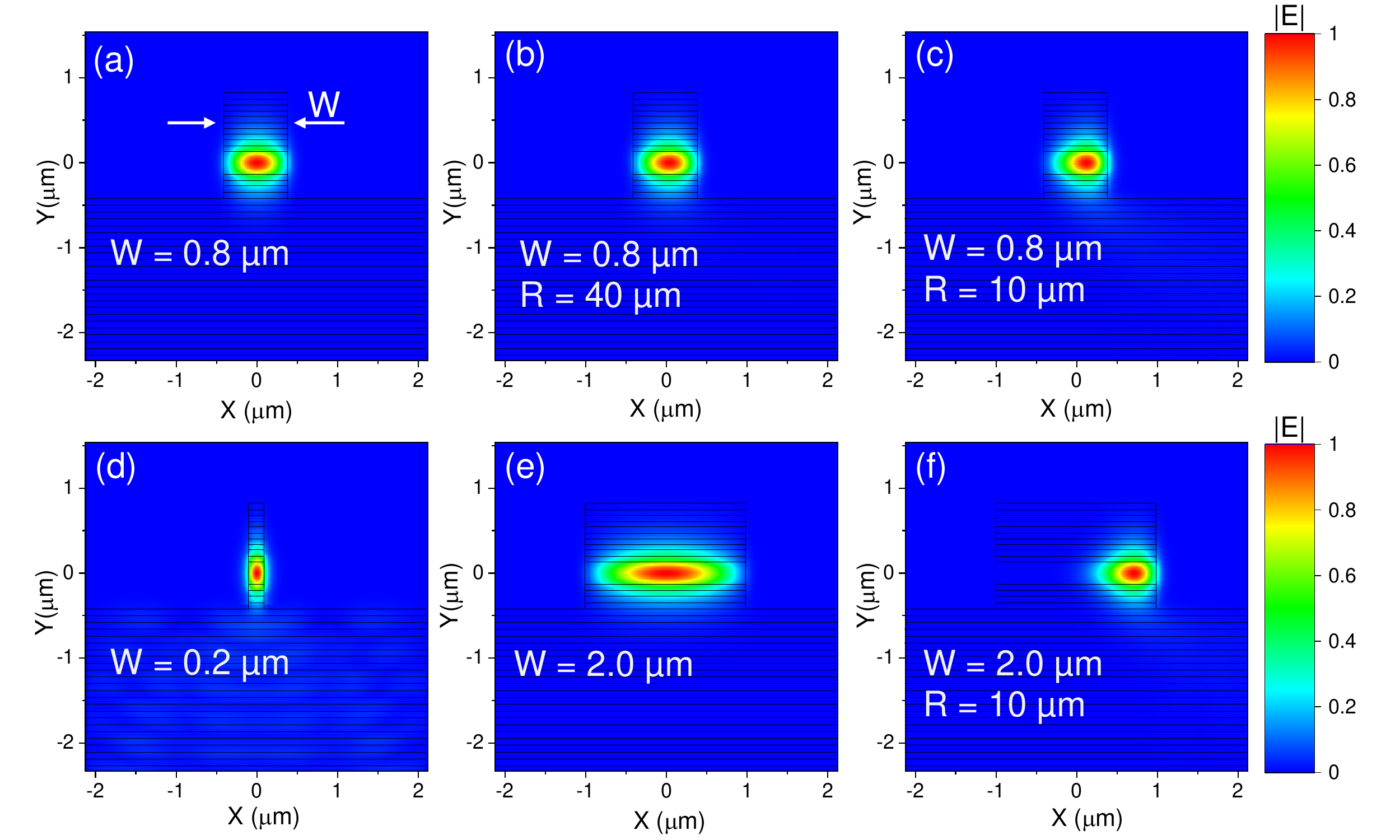}
		\caption{a)~Optical power distribution profile for the fundamental TE mode obtained in the case of 0.8~$\mu$m width (a)~bus WG, (b)~40~$\mu$m radius ring WG and (c)~10~$\mu$m radius ring WG; (d)~0.2~$\mu$m width bus WG, 2.0~$\mu$m width (e)~bus WG and (f)~10~$\mu$m radius ring WG. All modes are calculated at 930~nm wavelength.
		}
		\label{fig:modes}
	\end{figure}

	Figure~\ref{fig:modes} shows simulated optical mode profiles of the light field confined in our WG devices calculated for the transverse electric (TE) modes at 930 nm wavelength. In all cases, the modes are mainly concentrated within the GaAs cavity and partially penetrate the top and bottom DBR mirrors. For bent WGs optical field maximum is shifted towards ring outer edge.

	\newpage
	\section{Circular Bragg reflectors}
\begin{figure}[h]
		\centering
		\includegraphics[width=5in]{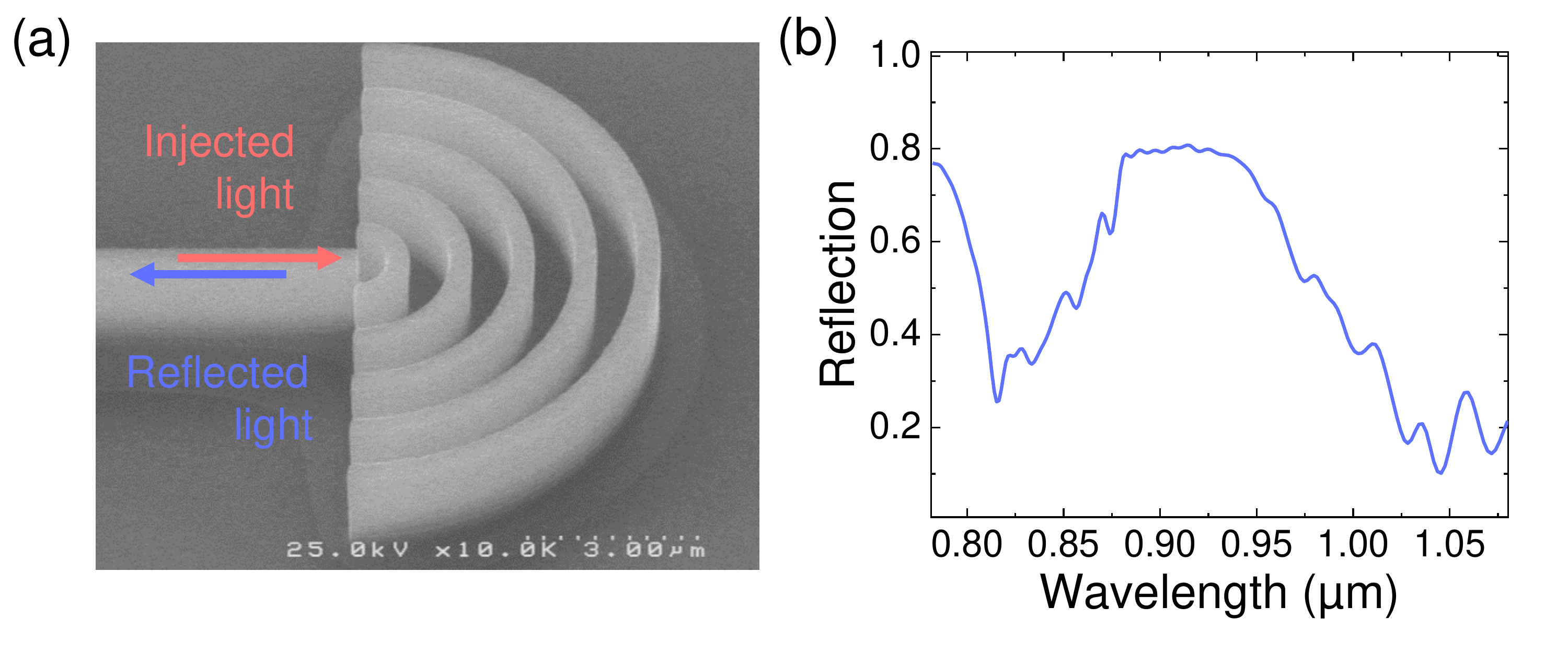}
		\caption{a)~Scanning electron microscope image of the fabricated circular Bragg reflector interconnected with a waveguide. b)~Simulated reflection spectra of the injected fundamental mode. The width of the air gaps is $728$\,nm and the width of the half rings is $357$\,nm. The device was optimized to achieve stop band centered around $930$\,nm. The maximum reflectivity is $80\%$.
		}
		\label{fig:mirrors}
	\end{figure}

	To effectively collect all the photons coupled from the ring into the single arm of the bus WG, we fabricated circular Bragg gratings. Purpose of those structures was to inverse the direction of light propagation in the WG. Circular Bragg design instead of the standard rectangular grating was chosen to limit scattering and collect light exiting WG at large angles. In Figure~\ref{fig:mirrors}, the Lumerical FDTD simulation of the reflection spectra for the Bragg grating with the air gaps of $728$\,nm and the width of the half rings $357$\,nm is shown. This device was optimized for maximum reflection at the center wavelength of $930$\,nm. Reflection on the level of 80\% is expected in the 875-950~nm range. Higher reflectivity could be potentially achieved by further shape optimization (better fit to the WG numerical aperture).
	
	\newpage
	\section{Inverse taper out-couplers}
\begin{figure}[h]
		\centering
		\includegraphics[width=1\linewidth]{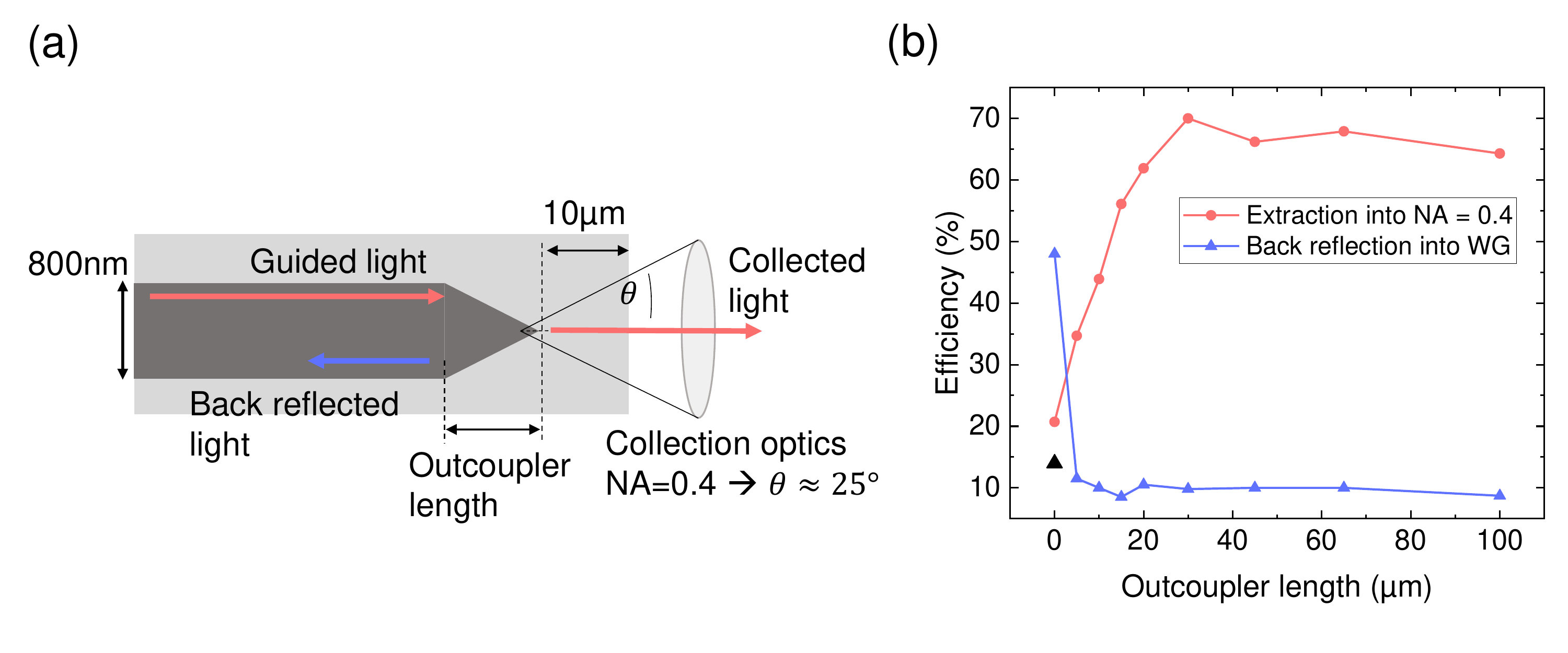}
		\caption{a) Scheme of the simulated outcoupler. b) Simulated coupling efficiency into the collection objective and the back-reflection into the waveguide. The optimization was done for a taper ending $10\,\mu$m away from the cleaved edge. For comparison the coupling efficiency of an cleaved waveguide ($d=0\,\mu$m and outcoupler length $=0\,\mu$m) is shown.}
		\label{fig:outcouplers}
	\end{figure}

	To efficiently extract photons from the waveguide into the off-chip collection optics tapered out-couplers are used. The out-coupler is designed in the form of a short inverse taper and shown schematically in Fig.~\ref{fig:outcouplers}a. In order to get side access to the circuit, a sample is cleaved around 10~$\mu$m from the out-couplers edge. The geometry of the taper was optimized through the 3D FDTD calculation. For that, the fundamental mode was injected into DBR ridge WG and then the far-field projection of the emission integrated over the acceptance angle of the NA$=0.4$ collection optics. Additionally, the fraction of WG back-reflected light was monitored. The length of the out-coupler was increased progressively to minimize the residual reflection and optimize collection efficiency (see Fig.~\ref{fig:outcouplers}b). For the out-coupler length of $30\,\mu$m placed 10~$\mu$m from the edge, collection efficiency of 70\% can be achieved with 0.4 NA, and reflection minimized to 10\%. In the case of non-terminated WG (cleaved WG facet), the coupling efficiency is limited to around 15\%, while reflection is close to 50\%.
	
	\newpage	
	\section{Waveguide mode converters}
\begin{figure}[h]
		\centering
		\includegraphics[width=6in]{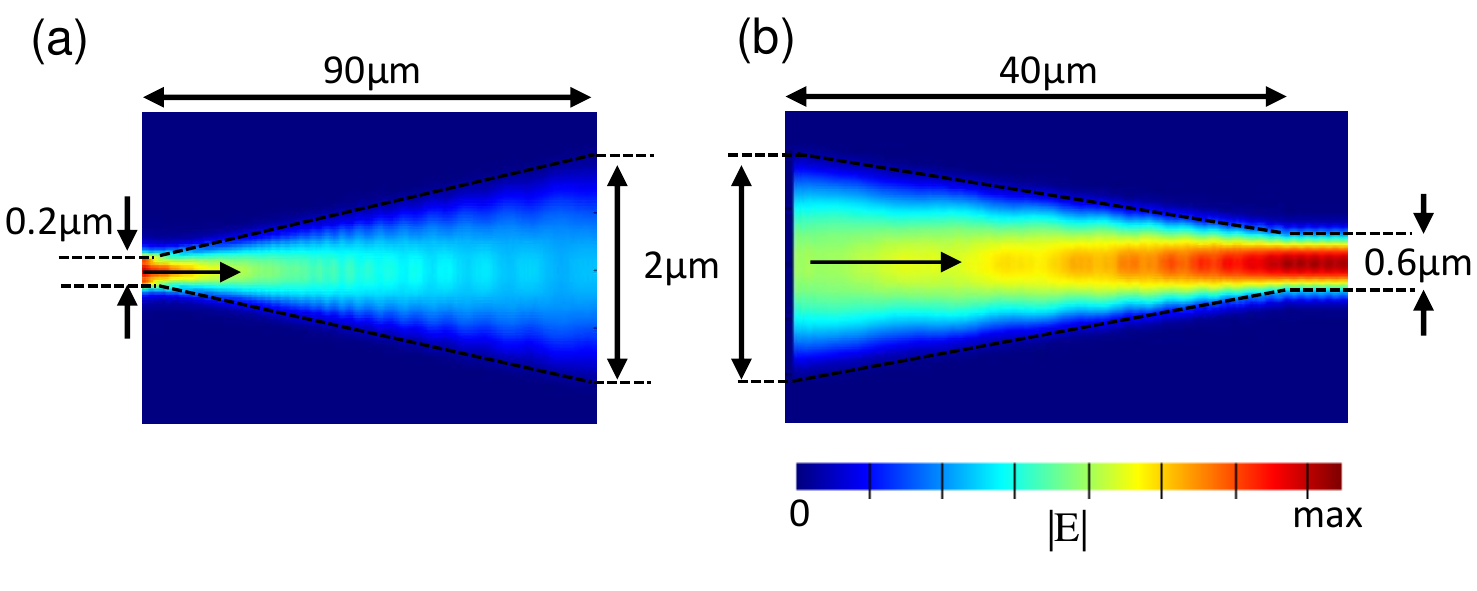}
		\caption{Waveguide mode converters. a) Optical intensity field distribution of the single-mode to the multi-mode converter. The fundamental mode is injected in a $0.2\,\mu$m waveguide tapered to a $2\,\mu$m waveguide, preserving $97\%$ of the transmitted light in the fundamental mode with overall $80\%$ transmission efficiency. b) Optical intensity field distribution of the multi-mode to the single-mode converter. Injecting the light in the fundamental mode of a $2\,\mu$m waveguide tapered to $0.6\,\mu$m WG preserves $99\%$ of transmitted light in the fundamental mode with total $97\%$ transmission efficiency.}
		\label{fig:tapers}
	\end{figure}

	Within our devices, we use both single-mode (SM) and multi-mode (MM) waveguides. In order to efficiently couple light between them, we use mode converters in the form of tapering sections. In the case of coupling light from SM to MM waveguide, fundamental mode conversion into the higher-order modes can be circumvented by choosing a sufficiently long tapering section preserving light guiding in the fundamental mode over the entire MM WG length. The required tapering lengths and corresponding performances are summarized in figure~\ref{fig:tapers}. The simulations were performed by 3D FDTD Lumerical Solutions and the mode occupation was calculated by expanding the light field after the tapering section using the mode expansion technique.
	
	\newpage
	\section{Ring resonator mode volume vs WG width}
\begin{figure}[h]
		\centering
		\includegraphics[width=3.5in]{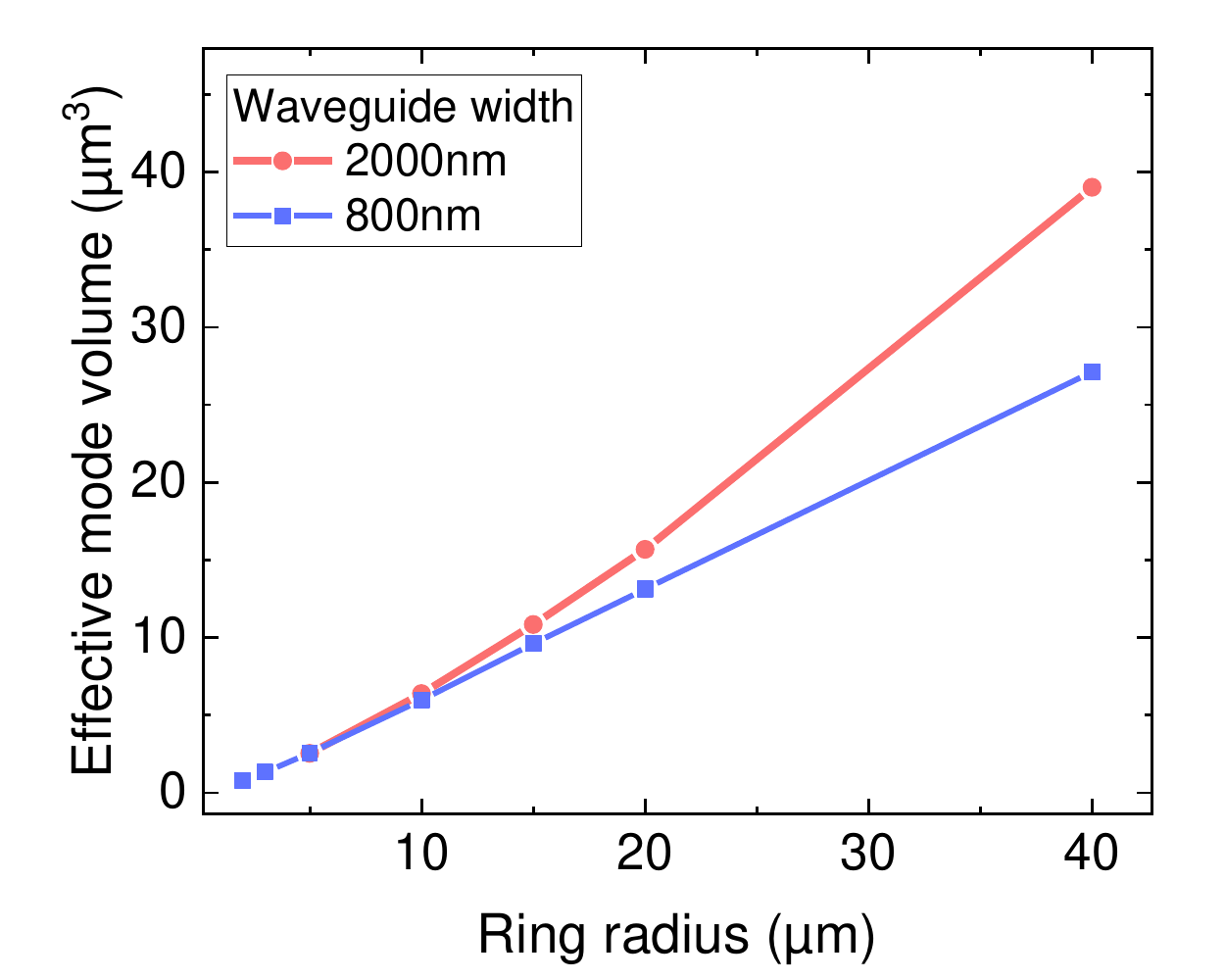}
		\caption{Effective mode volume vs ring radius calculated for different waveguide widths.}
		\label{fig:modeArea_and_VeffvsR}
	\end{figure}

	To verify how does the width of the WG influence the Purcell factor, the effective mode volume of the ring cavity was calculated. Results of simulations are summarized in Figure~\ref{fig:modeArea_and_VeffvsR}. The effective mode volume does not change significantly for ring radius smaller than 20~$\mu$m for WG width in the range of $0.8\,\mu$m to $2\,\mu$m. The effective ring mode volume was calculated by Lumerical FDTD, based on fundamental 2D WG mode profiles.
	
	\newpage
	\section{Ring design optimization}
	To optimize the design of the ring resonators for maximized Purcell we first consider the well-known formula for Purcell factor
	$
	F_p = \frac{3}{4\pi^2} \left( \frac{\lambda_{cav}}{n} \right) ^3 \frac{Q}{V},
	$
	for the case where the dipole is resonant with the cavity and also ideally positioned and oriented with respect to the local electric field, where $\lambda$ is the wavelength, $n$ is the refractive index, Q is quality factor and V mode volume of the cavity mode. Following on above, optimization of the structure can be achieved upon maximizing the $Q/V$ ratio. In ring structures, Q factor can be varied by controlling the gap distance between ring and bus WG, and mode volume via controlling the ring diameter and the WG width. Following on that optimal design would correspond to small diameter rings with relatively large gaps. However, this simple picture breaks down if the WG bending losses and sidewalls imperfection are also considered. To take those effects into account we performed two-steps optimization. 
	\begin{figure*}[h]
		\centering
		\includegraphics[width=6.0in]{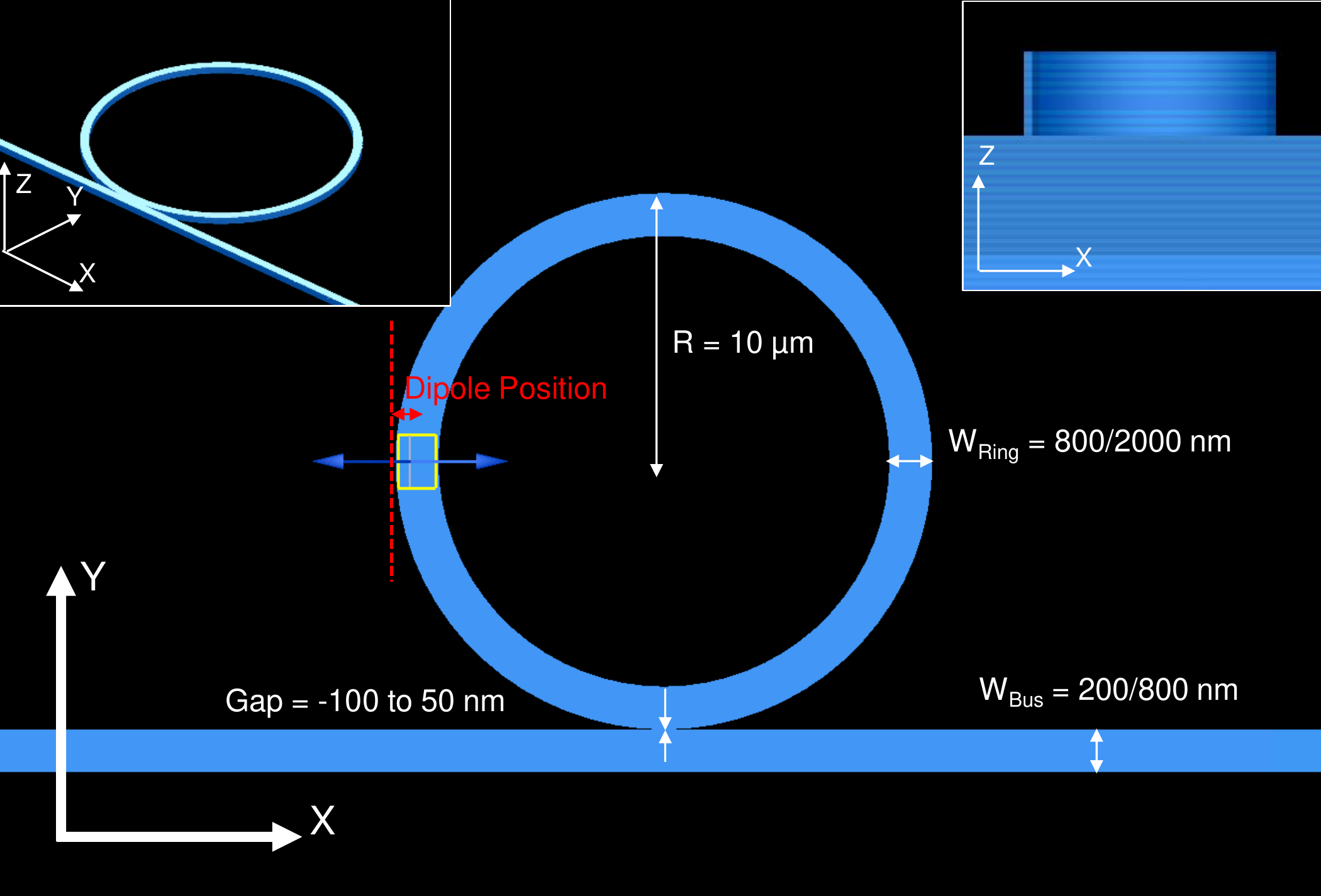}
		\caption{Scheme of the ring structure with labeled important parameters.}
		\label{fig:model}
	\end{figure*}
	As a first step, we experimentally check Q-factors obtainable for given ring diameter and gap size in our system. For that, we fabricated a set of devices with various radius and ring-bus-WG spacings and performed a systematic check of the Q-factors via optical measurements (PL at high pumping power). We noticed that for 5~$\mu$m radius rings modes with Q factors below 2k are usually observed [see Fig.~\ref{fig:Qfactor}(a)], while for 3~$\mu$m radius rings, cavity modes could not be resolved at all.
	\begin{figure}[t]
		\centering
		\includegraphics[width=6.6in]{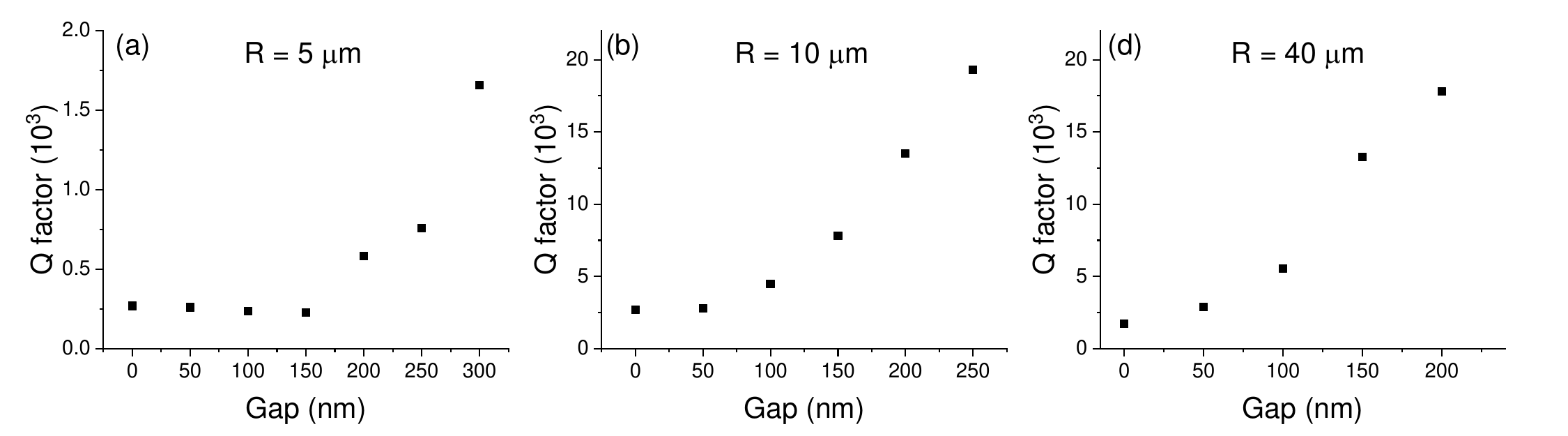}
		\caption{Experimentally obtained quality factors of the ring cavity modes in the function of the nominal gap distance between ring and bus WG.}
		\label{fig:Qfactor}
	\end{figure}
	For 10~$\mu$m radius rings, modes with much better quality factors are observed, Q of 2.5-3k for gap-less structures and Q up to 18k for 250~nm nominal gap [see Fig.~\ref{fig:Qfactor}(b)]. Similar results are achieved for larger diameter rings [see Fig.~\ref{fig:Qfactor}(c)]. An over tenfold improvement in the cavity line-width can be related to smaller bending losses and reduced light scattering on the outer ring sidewalls (mode localized more faraway from the defects). It is important to note here, that within our fabrication process gaps smaller than 100~nm could not be realized due to the e-beam proximity effect (for nominal 50~nm gap structures rings are in fact in contact with bus WG). Moreover, gaps designed as 100-120~nm are in some cases not perfectly etched through the GaAs cavity, which effectively decreases cavity quality factor (see SEM images in Fig.~\ref{fig:struc}). By considering the Q/V ratio for our structures, which in the case of rings could be simplified to Q/R it is clear that devices with the radius of 10~$\mu$m are the most promising for the Purcell enhancement purposes.
	\begin{figure}[h]
		\centering
		\includegraphics[width=5.2in]{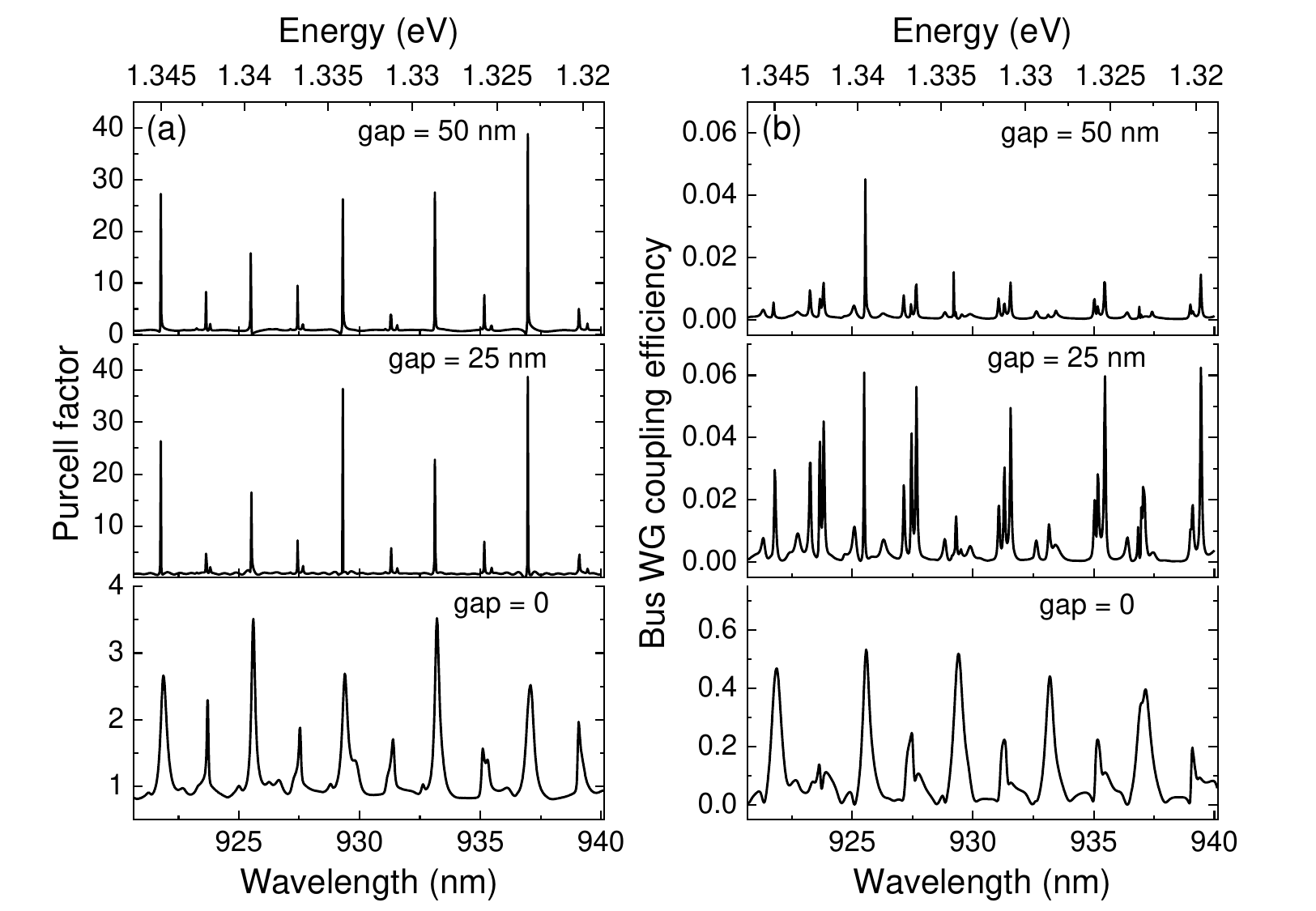}
		\caption{(a)~Purcell factor and (b)~WG coupling efficiency vs wavelength for ring-bus gap sizes of varying from to 0 to 50~nm. Simulations are performed for 10~$\mu$m radius rings with 2~$\mu$m width coupled to 0.2~$\mu$m bus WGs.}
		\label{fig:gapP}	
	\end{figure}
	\begin{figure}[h]
		\centering
		\includegraphics[width=5.0in]{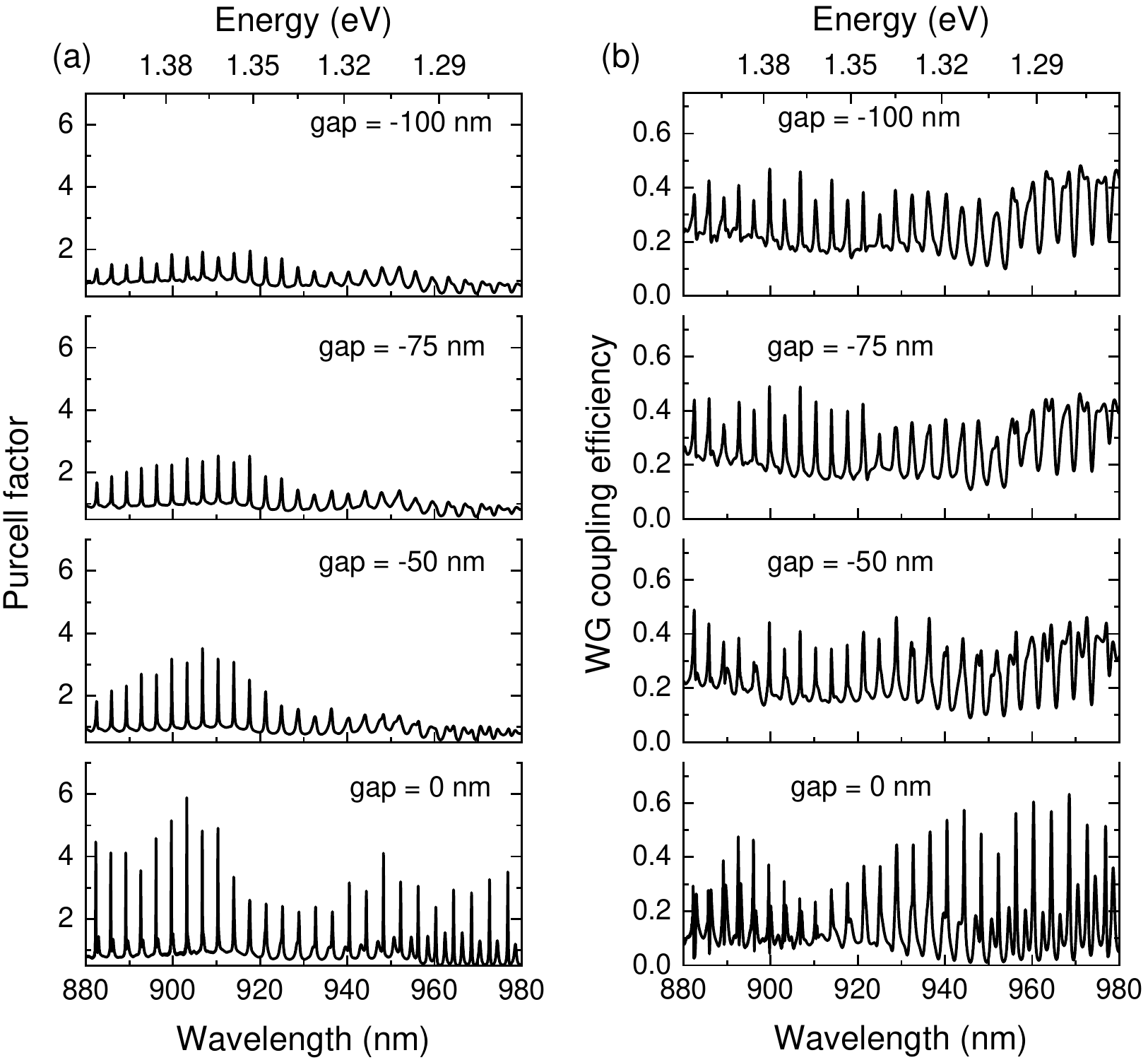}
		\caption{(a)~Purcell factor and (b)~WG coupling efficiency vs wavelength for $R$~=~10~$\mu$m ting structure with ring-bus-WG gap distance varying from of -100~nm to 0. (minus sign corresponds to ring placed inside bus WG). In order to clearly resolve the low Q fundamental radial modes, simulations are performed for 0.8~$\mu$m width rings.}
		\label{fig:gap}	
	\end{figure}
	
	Next, for the 10~$\mu$m radius rings full 3D FDTD simulations were performed for various gaps and dipole positions focused on Purcell factor and extraction efficiency. In Fig.~\ref{fig:model} scheme of the ring structure with labeled parameters of merit is shown. We note here, that FDTD simulations for rings as large as 10~$\mu$m are very demanding computationally thus rather limited parameter space was checked. First, we analyzed how does the gap between ring and bus WG influence the Purcell and bus WG coupling efficiency. In the case of the ring with 2~$\mu$m width coupled to 0.2~$\mu$m width bus WG we varied gap distance from 0 to 50~nm in 25~nm steps as shown in Fig.~\ref{fig:gapP}. Change of the gap distance from zero to 25~nm strongly increases both the Q and Purcell by around an order of magnitude. Further increase of the gap up to 50~nm does not affect Purcell anymore, suggesting that Q factor is already saturated for the 25~nm gap case. Even though high Purcell is expected for 25-50~nm gap rings, FDTD simulations predict relatively low coupling into the bus WG on the level of a few percent. We speculate that low coupling into bus WG is related to weak evanescent wave-like coupling between the ring and bus WG in respect to the losses into free space via bending. Further shortening of the gap distance or redesign of the coupling region into pulley configuration might be advantageous in this respect.   
	
	\begin{figure*}[h]
		\centering
		\includegraphics[width=6.8in]{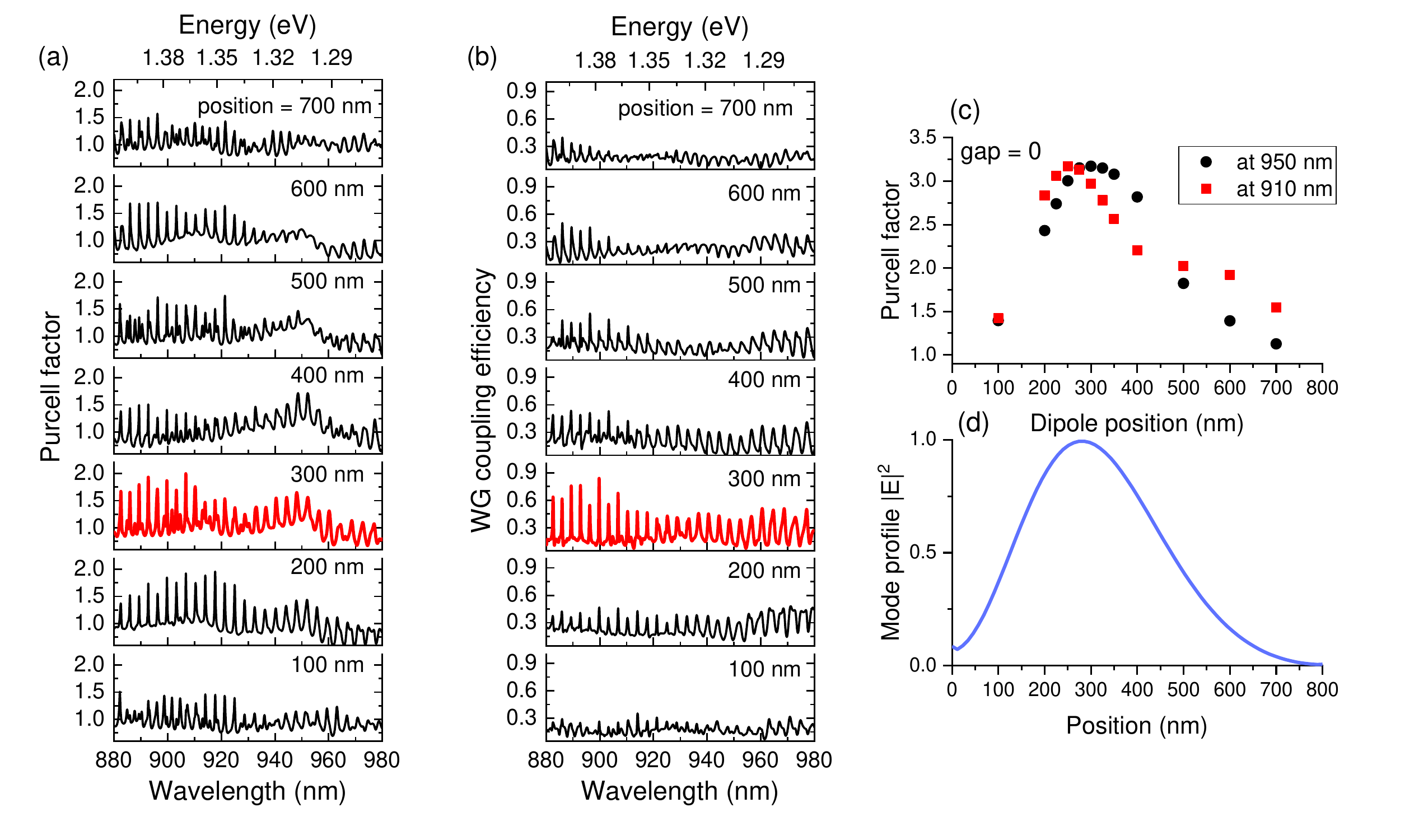}
		\caption{(a)~Purcell factor and (b)~coupling efficiency into the bus WG vs wavelength for 10~$\mu$m radius ring placed 100~nm inside the bus WG (gap -100~nm) for dipole positioned 100-700~nm from the outer edge of the ring. (c)~Purcell factor vs dipole position for 10~$\mu$m radius gap-less ring mode at 950~nm and 910~nm wavelength. (d)~Bent WG fundamental mode profile at 930~nm wavelength. Simulations are performed for 0.8~$\mu$m width WGs.}
		\label{fig:position}
	\end{figure*}
	
	Next, for 0.8~$\mu$m width rings coupled to 0.8~$\mu$m width bus WG, we varied the gap distance from -100 nm up to 0, where negative values correspond to the ring placed inside the bus WG. As can be seen in Fig.~\ref{fig:gap}, placing a ring within the bus WG significantly reduces the Purcell factor, while the WG coupling efficiency is only slightly modified. In fact, by optimizing the dipole position for the -100 nm gap ring, coupling efficiency as high as 0.88 is possible with rather moderate Purcell of around 2, as shown in Fig.~\ref{fig:position}(a-b). We speculate that further optimization of the ring-bus-WG coupler for instance by using pulley design might allow for achieving even higher coupling efficiency. Similar dipole position optimization routine has been repeated for the gap-less ring design. In Fig.~\ref{fig:position}(c) Purcell factor vs dipole position is plotted for two modes at wavelengths close to 950 and 910~nm. Maximal Purcell is obtainable for dipole placed around 200-300~nm from the outer edge of the ring and slightly depend on the wavelength following the ring mode profile [see Fig.~\ref{fig:position}(d)].
	
	Unfortunately full 3D FDTD simulations for the R~=~40~$\mu$m rings were not possible due to the computational memory constraints. To get an estimate of the possible Purcell in the case of R~=~40~$\mu$m devices we extrapolate the R~=~10~$\mu$m ring results by decrease of the Purcell by 4, related to the 4-times larger mode volume. In such case maximal Purcell of 10 is expected for 100k quality factor modes, Purcell of 2 for 20k Q and no-Purcell for Q$\leq$10k.

	\newpage	
	\section{Waveguide transmission losses}
\begin{figure}[h]
		\centering
		\includegraphics[width=6.5in]{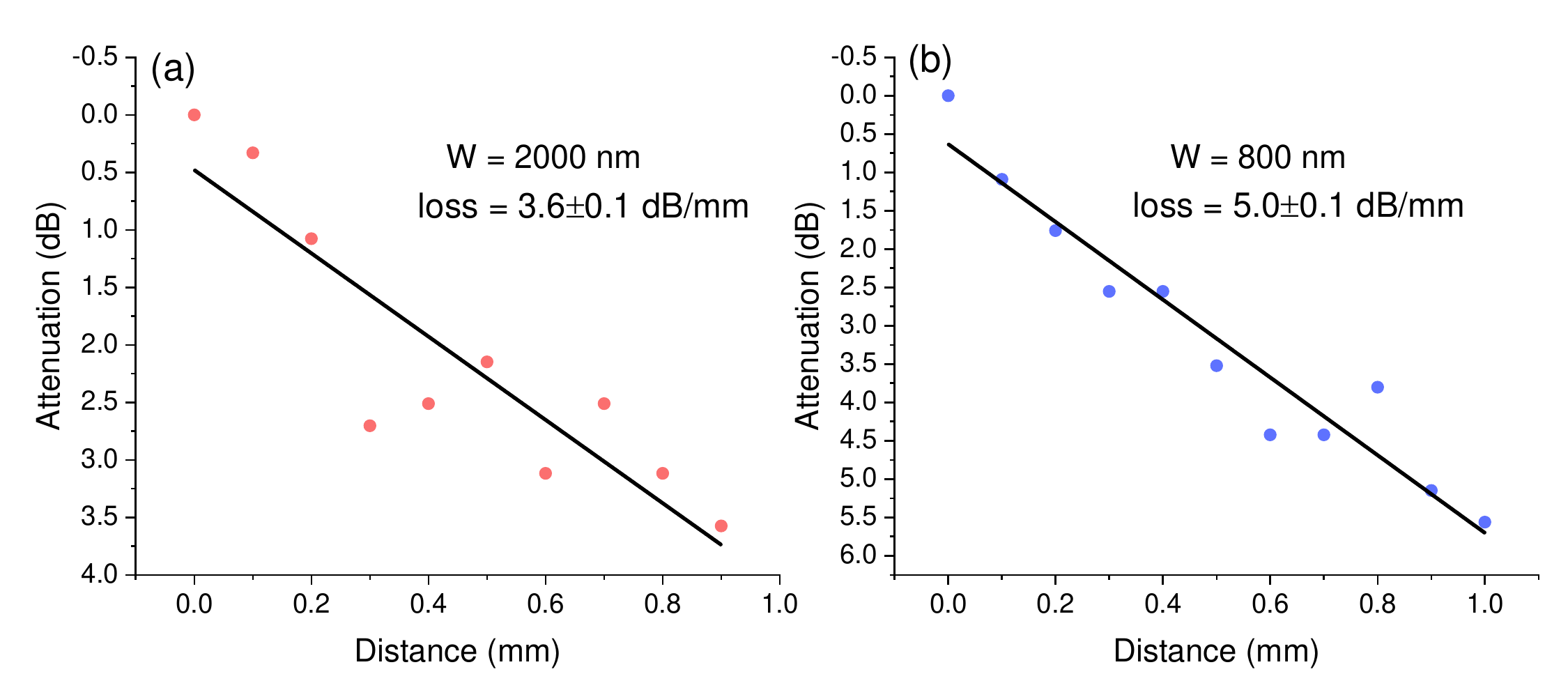}
		\caption{Waveguide transmission losses vs distance from outcoupler for (a)~2000~nm width MM bus waveguide, (b)~800~nm width SM bus waveguide.}
		\label{fig:loss}
	\end{figure}

	To estimate the structural quality of the fabricated ridge waveguides, the optical WG transmission losses were determined. For that purpose sample was excited with very high pumping power, allowing to observe spectrally broad QD ensemble emission. The beam spot was scanned along the bus WG and the QD ensemble emission detected from the side. Figure~\ref{fig:loss}a and b show the corresponding attenuation of the measured intensities at 915~nm plotted as a function of the distance to the out-couplers. Selected ridges exhibit transmission losses on the level of 3.6$\pm$0.1~dB/mm for 2000~nm width WG and 5.0$\pm$0.1~dB/mm for 800~nm WG. Waveguide transmission is limited by the ridge sidewalls imperfections, which could be potentially further improved by optimizing the fabrication process.
	
	\newpage	
	\section{Extraction efficiency estimation}
	In order to estimate the total device extraction efficiency as well as on-chip QD-WG coupling efficiency, the setup transmission characteristics was carefully checked with cw laser line tuned to around 927~nm wavelength of interest. Since the far-field emission pattern from the tapered out-coupler was not optimized for the optical fiber coupling, we performed a calibration based on the number of integrated counts observed on the CCD camera vs the number of counts observed on the SSPD (quantum efficiency of 55\% estimated by measuring a laser signal of known power for which we can deduce the number of incident photons) at 927~nm. For that, the laser was coupled into a single-mode optical fiber and then introduced (i) directly into SSPD detector or, (ii) monochromator combined with CCD. Additionally, the total transmission of side detection setup was checked (39.7\% transmission from inside of cryostat to monochromator entrance). In result, we derived a calibration constant, allowing us to estimate the total single-photon rate of 1.8MHz at 76MHz driving extracted into first collection optics, which corresponds to the extraction efficiency of 2.35\%. For further calculations we assume WG transmission of T$_{WG}$~=~0.5 (losses of 3dB/mm), the out-coupler transmission of T$_{out}$~=~0.7, Bragg reflector efficiency of R$_{ref}$~=~0.8, and the 0.2~$\mu$m to 2.0~$\mu$m taper transmission of T$_{tap}$=0.8. The total device transmission in such case can be quantified as: T$_{tot}$~=~(0.5T$_{tap}$+ 0.5T$_{tap}\cdot$R$_{ref}\cdot$T$_{tap}$)$\cdot$T$_{WG}\cdot$T$_{out}$. In the case of the perfect device (maximal transmission values are taken from FDTD simulations), T$_{tot}$ would correspond to 0.23, which allows estimating the lower limit of QD-WG coupling efficiency on the level of 10.3\%. It needs to be noted that this value is most likely strongly underestimated, as the perfect performance of each functionality is assumed. This concerns especially the out-coupler which is the most sensitive to fabrication imperfections. In the case of taper out-couplers, we observe usually an factor of 3 improvement in the collection in respect to non-tapered WGs, which again assuming perfect collection from cleaved facet T$_{cle}$ for 2.0~$\mu$m width WG of 0.10, corresponds to 0.3 transmission T$_{out}'$ in the case of tapered out-coupler. Assuming this more realistic value, the extraction efficiency of 24\% can be expected.  
	
	\newpage
	\section{HOM histogram fitting and visibility correction}
	Experimental HOM histogram data, have been fitted with the two-sided exponential decay function convoluted with IRF with a central peak intensity ratio of A$_{-6ns}$:A$_{-3ns}$:A$_0$:A$_{+3ns}$:A$_{+6ns}$
	\begin{equation}
	\begin{split}
	F(t) = & X \sum_{\pm} \left(A_{0} e^{\frac{-\left|t\right|}{\tau_{dec}}}
	+ A_{\pm 3ns} e^{\frac{-\left|t\pm\tau_{pd}\right|}{\tau_{dec}}}
	+ A_{\pm 6ns} e^{\frac{-\left|t\pm 2\tau_{pd}\right|}{\tau_{dec}}} \right) \\
	+ X \sum_{k,\pm} & \left( 3(A_{-6ns}+A_{+6ns}) e^{\frac{-\left|t\pm k\tau_{rr}\right|}{\tau_{dec}}}
	+ (3A_{\pm 6ns}+A_{\mp 6ns}) e^{\frac{-\left|t\pm k\tau_{rr} \pm \tau_{pd} \right|}{\tau_{dec}}}
	+ A_{\pm 6ns} e^{\frac{-\left|t\pm k\tau_{rr} \pm 2\tau_{pd} \right|}{\tau_{dec}}} \right),
	\end{split}
	\end{equation}
	where $A_0$ is corresponding to the zero delay peak amplitude, $X=\frac{2A}{A_{-3ns} + A_{+3ns}}$ is normalization constant, A corresponds to the mean value of the $\pm3ns$ peaks amplitude, $\tau_{dec}$ is a resonance fluorescence decay time, $\tau_{pd}$ is a delay time between the pulse trains (3~ns), $\tau_{rr}$ is a repetition time of the laser (13.16~ns). The first line of the equation 1 corresponds to the central cluster of peaks, while the second is the sum over non-central clusters. In the case of perfectly balanced HOM interferometer and ideal single-photon source 1:2:0:2:1 ratio for central, and 1:4:6:4:1 ratio for non-central peaks is expected. In the case of interferometer with non-perfect beam-splitters reflection/transmission ratio (R/T)
	\begin{equation}
	A_{-6ns} = TR^3,
	A_{+6ns} = RT^3,
	A_{-3ns} = A_{+3ns} = TR^3+RT^3,
	\end{equation}
	which introduces slight imbalance between $\pm$6~ns peaks. For the fitting procedure all values beside the $A_0$ and $A$ (fitting parameters), were fixed. The visibility has been calculated based on the central peak areas at zero and $\pm$3~ns delay, which have been directly obtained from the fitting parameter $A_0$.
	\begin{equation}
	V_{exp} = 1-\frac{2A_0}{A_{-3ns} + A_{+3ns}}.
	\end{equation}
	We note here that inclusion of the R/T ratio in the function F(t) in current form allows for achieving more precise fit but obtained visibility is still affected by the interferometer imbalance. To take into account influence HOM interferometer imperfections as well as non-zero $g^{(2)}(0)$ source extend, we corrected $V_{exp}$ value according to
	\begin{equation}
	V = V_{exp}[1+2g^{(2)}(0)]\cdot\frac{R/T+(R/T)^{-1}}{2(1-\varepsilon)^2},
	\end{equation}
	where $R/T$ is the beam-splitter reflectivity/transmission ratio and $(1-\varepsilon)$ is classical contrast of the HOM interferometer. In table~\ref{tab:table} raw and corrected values of the extracted visibilities are summarized. 
	
	
	\begin{table*}[h]
		\caption{\label{tab:table}Summarized values of recorded single photon purity and indistinguishability for each device.}
		\begin{tabular}{cccccccc}
			QD no.&R&Energy&$R/T$&$1-\varepsilon$&$V_{raw}$&$V$&$g^{(2)}(0)$\\ \hline
			1&10~$\mu$m&1.3370~eV&1.05&0.995&0.90&0.95$\pm$0.02&0.0191$\pm$0.0007 \\
			2&10~$\mu$m&1.3206~eV&1.38&0.990&0.81&0.93$\pm$0.02&0.035$\pm$0.002 \\
			3&40~$\mu$m&1.321~eV&1.16&0.990&0.85&0.95$\pm$0.02&0.040$\pm$0.002 \\
			
		\end{tabular}
	\end{table*}

	\newpage	
	\section{Performance of other ring resonator devices}
	Resonance fluorescence studies, similar to performed for the device no.1 described in the main text, have been also performed for two other ring resonators.
	
	\begin{figure}[!h]
		\includegraphics[width=5in]{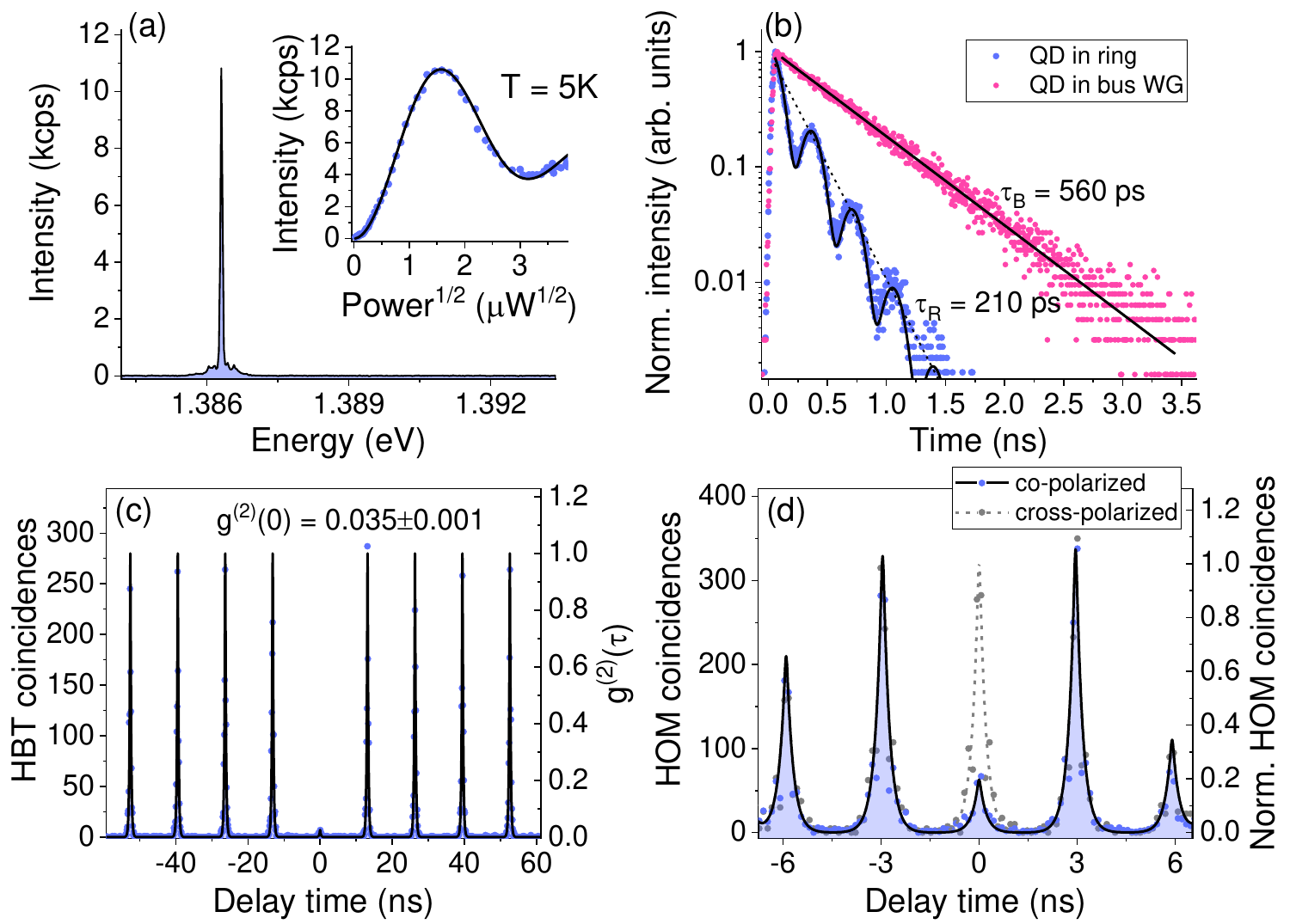}
		\caption{\label{fig:R10} Optical properties of QD$_2$ coupled to ring with radius $R$~=~10~$\mu$m. (a)~Pulsed resonance fluorescence spectrum of QD$_2$ at 4.5~K. Inset:~Resonance fluorescence intensity vs square root of power demonstrating Rabi oscillation. (b)~Time-resolved resonance fluorescence trace for QD perfectly tuned into cavity resonance and reference QD in bus WG. (c)~Second order correlation function histogram recorded under pulsed resonance fluorescence and $\pi$-pulse excitation power. (d)~Two-photon interference Hong-Ou-Mandel histogram recorded for 3~ns time separated co- (blue points) and cross-polarized (grey points) single photons under pulsed resonance fluorescence and $\pi$-pulse excitation power.}
	\end{figure}	
	
	Figure~\ref{fig:R10}(a) shows side detected pulsed resonance fluorescence spectra from QD$_2$ coupled the MM ring resonator with 10~$\mu$m radius and 100~nm nominal gap between the ring and tapered bus WG. This specific resonator can be characterized with a rather low-Q factor of 2k. We relate this with the non-perfectly etched gap between ring and bus WG and ring sidewalls imperfections. In Inset of Fig.~\ref{fig:R10}(c) the peak intensity versus the square root of the incident power is shown. Clear oscillatory behaviour with damping at higher power is observed, which is clear evidence of Rabi oscillations related to coherent control of the QD two-level system. Time-resolved resonance fluorescence measurements for QD under investigation and reference QD in bus WG is shown in Fig.~\ref{fig:R10}(b). In the case of QD coupled to ring, emission decay trace exhibit oscillations due to quantum beating between the two fine structure components of the exciton state. Data is fit by the function being a product of mono-exponential decay and sine function, with a decay time of 210~ps and the oscillation period of 345~ps corresponding to 2.9~GHz splitting of the fine structure. Due to the low-Q factor of the cavity, temperature spectral detuning of QD emission in this case shown only slight changes of the decay time from 210~ps to 300~ps for over 0.8~meV QD-cavity detuning. In order to extract Purcell factor direct comparison with emitters at the same energy but localized in the bus WG was performed, revelling enhancement of around 2.7, thus Purcell of 1.7. Next, HBT and HOM experiments were performed on resonance fluorescence signal for QD$_2$ under $\pi$-pulse excitation. In Fig.~\ref{fig:R10}(c) a second-order correlation function histogram is shown for QD$_2$ revealing $g^{(2)}(0)$ of $0.035\pm0.001$. In Fig.~\ref{fig:R10}(d) a two-photon interference histogram of QD$_2$ emission for orthogonal (gray dots) and parallel (black dots) polarized photons is presented. Upon fitting procedure, we obtain a raw two-photon interference visibility of 0.81$\pm$0.02 and corrected value of 0.93$\pm$0.02 (for more details see Tab.~\ref{tab:table}). 
	
\begin{figure}[!h]
		\includegraphics[width=6in]{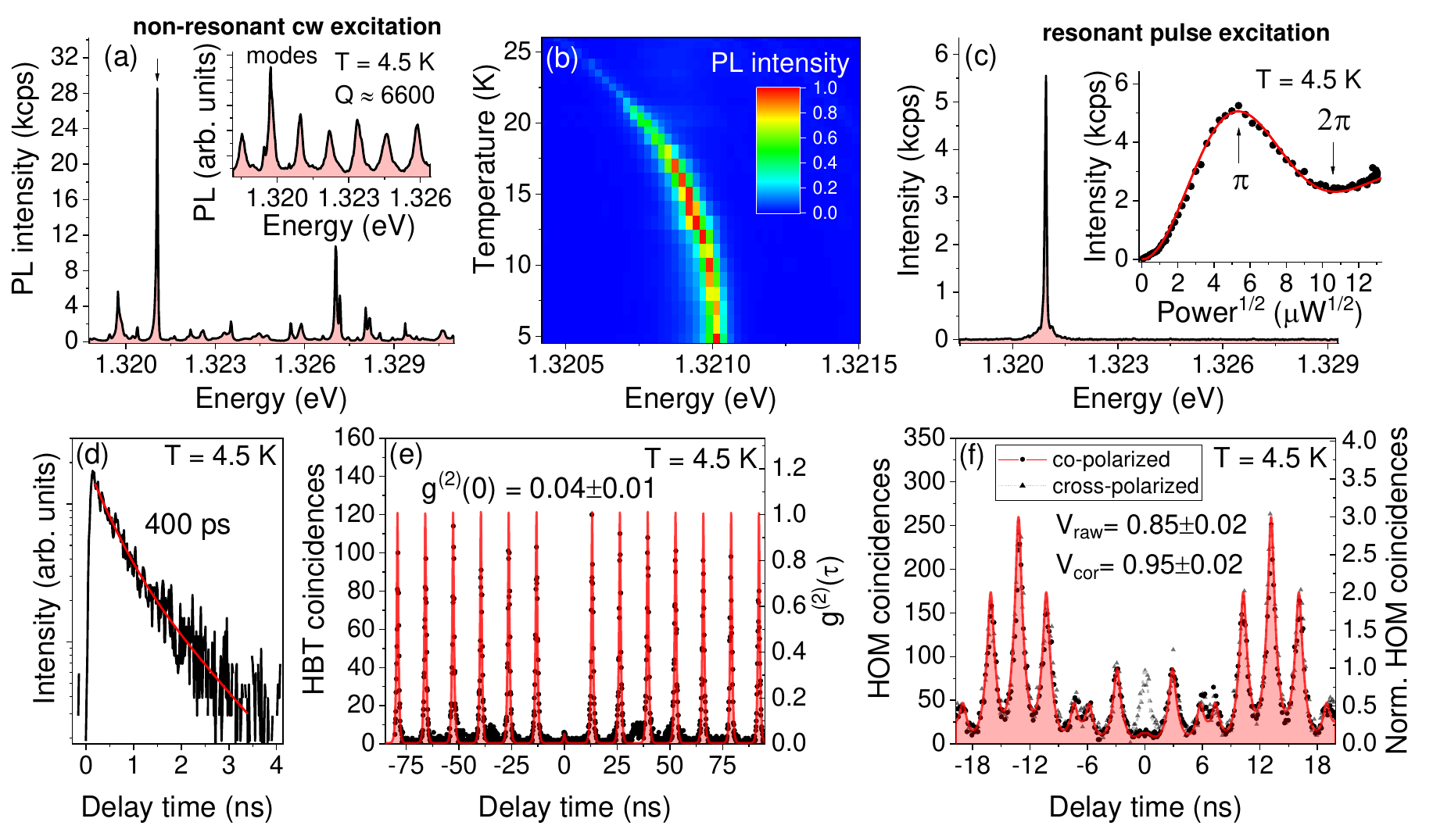}
		\caption{\label{fig:R40} Optical properties of QD$_3$ coupled to ring with radius $R$~=~40~$\mu$m. (a)~Side detected photoluminescence spectrum from QD$_3$ recorded under non-resonant cw excitation at a sample temperature of 4.5~K. Inset:~Optical modes of the ring under investigation. (b)~Temperature dependence of the QD$_3$ PL spectrum. (c)~Pulsed resonance fluorescence spectrum of QD$_3$ at 4.5~K. Inset:~Resonance fluorescence intensity vs square root of power demonstrating Rabi oscillation. (d)~Time-resolved resonance fluorescence trace for QD perfectly tuned into ring resonance. (e)~Second order correlation function histogram recorded under pulsed resonance fluorescence and $\pi$-pulse excitation power. (f)~Two-photon interference Hong-Ou-Mandel histogram recorded for 3~ns time separated co- (black points) and cross-polarized (grey points) single photons under pulsed resonance fluorescence and $\pi$-pulse excitation power.}
	\end{figure}

	Analogues studies were repeated for a selected 40~$\mu$m radius ring. Figure~\ref{fig:R40}(a) shows a side collected PL spectra from QD$_3$ under above-band gap cw excitation, with an emission line at 1.321~eV coupled to 0.8~$\mu$m width WG ring with 40~$\mu$m radius placed 100~nm within the single-mode bus WG (gap~=~-100~nm). The inset in Fig.\ref{fig:R40}(a) shows ring resonator modes for the investigated device with Q factors of around 6.6k. Similarly as in the case of QD$_1$, by tuning the QD$_3$ emission energy with temperature across the ring resonances a significant change of the PL intensity can be observed [see Fig.~\ref{fig:R40}(b)]. Figure~\ref{fig:R40}(c) shows side detected pulsed resonance fluorescence spectra from QD$_3$ at a temperature of 4.5~K. Power depended experiments shown in Fig.\ref{fig:R40}(c) reveal clear Rabi oscillations. In the case of a 40~$\mu$m radius ring with Q factor smaller than 7k, we expect rather negligible Purcell effect. This is partially confirmed by the time-resolved resonance fluorescence [see Fig.\ref{fig:R40}(d)] revealing decay time of around 400~ps at 4.5~K, corresponding to less than 1.3 enhancement in respect to reference 500~ps decay time. Next, we performed HBT and HOM experiments on resonance fluorescence signal for QD$_3$ under $\pi$-pulse excitation. In Fig.~\ref{fig:R40}(e) a second-order correlation function histogram is shown for QD$_3$ revealing $g^{(2)}(0)$ of $0.04\pm0.01$. In Fig.~\ref{fig:R40}(e) a two-photon interference histogram of QD$_3$ emission for orthogonal (gray dots) and parallel (black dots) polarized photons is presented. Upon fitting procedure, we obtain a raw value of two-photon interference visibility of 0.85$\pm$0.02 and corrected value of 0.95$\pm$0.02.
	
